\documentclass[usenatbib]{mn2e}
\usepackage{hyperref}
\hypersetup{
     colorlinks=true,
     linkcolor=blue,
     citecolor=blue,
     filecolor=blue,
     urlcolor=blue,
 } 
\usepackage{apjfonts,amsfonts,amsmath,amssymb,bm,ctable,verbatim}

\newcommand{\gizmourl}{\href{http://www.tapir.caltech.edu/~phopkins/Site/GIZMO.html}{\url{http://www.tapir.caltech.edu/~phopkins/Site/GIZMO.html}}}
\newcommand{\acknowledgments}[1]{\begin{small}\section*{Acknowledgments}\end{small}{\noindent #1}\vspace{5pt}}

\title[Cloud Survival in the CGM]{On the Survival of Cool Clouds in the Circum-Galactic Medium}

\author[Li et al.]{
\parbox[t]{\textwidth}{Zhihui Li\thanks{E-mail:zhihui@caltech.edu}$^1$, Philip F.~Hopkins$^1$, Jonathan Squire$^2$, Cameron Hummels$^1$} \vspace*{4pt} \\
$^1$ TAPIR, Mailcode 350-17, California Institute of Technology, Pasadena, CA 91125, USA \\
$^2$ Department of Physics, University of Otago, 730 Cumberland St, North Dunedin, Dunedin 9016, New Zealand \\}

\date{}
\begin{document}
\maketitle

\begin{abstract}
We explore the survival of cool clouds in multi-phase circum-galactic media. We revisit the ``cloud crushing problem'' in a large survey of simulations including radiative cooling, self-shielding, self-gravity, magnetic fields, and anisotropic Braginskii conduction and viscosity (with saturation). We explore a wide range of parameters including cloud size, velocity, ambient temperature and density, as well as a variety of magnetic field configurations and cloud turbulence. We find that realistic magnetic fields and turbulence have weaker effects on cloud survival; the most important physics is radiative cooling and conduction. Self-gravity and self-shielding are important for clouds which are initially Jeans-unstable, but largely irrelevant otherwise. Non-self-gravitating, realistically magnetized clouds separate into four regimes: (1) At low column densities, clouds evaporate rapidly via conduction. (2) A ``failed pressure confinement'' regime, where the ambient hot gas cools too rapidly to provide pressure confinement for the cloud. (3) An ``infinitely long-lived'' regime, in which the cloud lifetime becomes longer than the cooling time of gas swept up in the leading bow shock, so the cloud begins to accrete and grow. (4) A ``classical cloud destruction'' regime, where clouds are eventually destroyed by instabilities. In the final regime, the cloud lifetime can exceed the naive cloud-crushing time owing to conduction-induced compression. However, small and/or slow-moving clouds can also evaporate more rapidly than the cloud-crushing time. We develop simple analytic models that explain the simulated cloud destruction times in this regime.
\end{abstract}


\section{Introduction}
\label{sec:intro}

The circum-galactic medium (CGM) is the diffuse, multi-phase gas surrounding a galaxy inside its virial radius and outside its disk and interstellar medium. In recent years, observations and simulations have revealed that CGM plays a significant role in galaxy evolution, in the sense that it both supplies gas for the galaxy's star formation and recycles the energy and metals produced by stellar and AGN feedback \citep{2017ARA&A..55..389T}. 

Over the past twenty years, direct observations have revealed the complex multi-phase structure in the CGM, in its ionization structure and dynamics. It is customary to classify the CGM gas into three components in different physical states \citep{2013ApJ...770..139C}, namely: (a) the cool gas phase ($T < 10^{5}$\,K), mainly composed of neutral hydrogen and low ionization-potential ions like Mg II, Si II and C II (e.g., \citealt{1996ApJ...471..164C, 1998ApJ...498...77C, 2010ApJ...717..289S, 2014ApJ...796..140P, 2017ApJ...850L..10J}); (b) the warm-hot gas phase ($T \sim 10^{5}-10^{6}$\,K), specifically the high ionization-potential ions like C III, C IV, O VI, and Ne VIII (e.g., \citealt{2006ApJ...641..217S, 2011ApJ...743..180S, 2014ApJ...792....8W}); (c) the hot gas phase ($T >$ $10^{6}$\,K), consisting even more highly ionized species, like O VII and O VIII (e.g., \citealt{2008SSRv..134...25R, 2010ApJ...716.1514Y}). Different ions in different physical states also display varied kinematics, resulting in a variety of absorption line profiles \citep{2016ApJ...833...54W}.

The existence of multi-phase gas raises fundamental questions about how the ``cool'' phases can be maintained. While the CGM can be thermally unstable, it is well-known from ideal-hydrodynamic simulations that a cool cloud moving through a hot medium  at any appreciable velocity will  be rapidly ``shredded'' and destroyed (mixed into the hot medium) by a combination of shocks, Rayleigh-Taylor, Kelvin-Helmholtz, and related instabilities \citep{1975ApJ...195..715M}. If clouds are ``ejected'' from the galaxy directly in a cool phase of galactic outflows, or form ``in-situ'' in outflow cooling shocks/shells, they are expected to have large (super-sonic) relative velocities to the ambient medium \citep{2016MNRAS.455.1830T}. Even if they form in-situ in a thermally-unstable hydrostatic CGM ``halo'' of hot gas around the galaxy, they are buoyantly unstable and will ``sink'' at trans-sonic velocities \citep{2018MNRAS.473.5407M}. 

The simple formulation of this problem -- namely the survival of a cold cloud moving through a hot ambient medium -- is the classical ``cloud crushing'' problem, and has been studied for several decades in the context of the interstellar medium (ISM), particularly for the case of giant molecular clouds (GMC) being hit by supernova shocks (e.g., \citealt{1977ApJ...211..135C, 1977ApJ...215..213M, Klein1994}). However, in the CGM, the dominant physics and their effects are expected to be very different from those in the ISM. For example, GMCs are marginally self-gravitating, highly supersonically-turbulent (turbulent Mach numbers $\sim 10-100$), molecular and  self-shielding\footnote{By ``self-shielding'' we mean the cloud column density is high enough to absorb all the incoming ionizing photons from the meta-galactic UV background and shield the inner neutral gas from being ionized.} (temperatures $\sim 10-1000\,$K, column densities $\gtrsim 100\,M_{\odot}\,{\rm pc^{-2}} \sim 10^{22}\,{\rm cm^{-2}}$), with ratios of thermal-to-magnetic pressure much less than one (plasma $\beta \ll 1$), and extremely short ion/electron mean-free-paths (negligible conduction/viscosity). CGM clouds, on the other hand, are generally not self-gravitating or Jeans-unstable, are ionized or atomic (non-molecular, non-self-shielded, with temperatures $\gtrsim 10^{4}\,$K), exhibit weakly sub-sonic or (at most) trans-sonic turbulence (turbulent Mach numbers $\lesssim 1$), and have dynamically negligible magnetic field strengths ($\beta \gg 1$). Further, given their lower densities and higher temperatures, such clouds can be comparable in size to the mean-free-paths of hot electrons in the ambient medium, meaning that conduction and viscosity could be extremely important. Moreover, those conduction/viscosity effects will be very anisotropic, given the small ratio of the particles' gyro radii to the system size, and could easily be in regimes where standard classical results break down.

All of this means that it is unclear how much, if any, intuition can be ``borrowed'' from the historical cloud-crushing studies in the ISM. As a result, there has been a recent resurgence of work on this idealized cloud-crushing problem but in the CGM context (e.g., \citealt{2015ApJ...805..158S, 2016ApJ...822...31B, 2016MNRAS.458.1164L, 2017MNRAS.470..114A, 2018arXiv180610688L, Gronke18, Gronke19, Sparre2019}). However, given the more recent nature of these studies and the computational expense of simulations including all of the physics above, this work has generally been limited in one of two ways: either (1) neglecting key physics (e.g.,\ ignoring radiative cooling, magnetic fields, anisotropic conduction/viscosity, saturation effects, or considering only two-dimensional cases), or (2) considering only a very limited parameter space (i.e., a couple of example clouds). In this paper, we therefore seek to build an analytical picture on the insights of these recent works by surveying an large parameter space of relevance to CGM clouds (e.g.,\ of cloud sizes, column densities, and velocities, as well as ambient temperatures, densities, and magnetic field properties). We include radiative cooling, magnetic fields, and fully-anisotropic conduction and viscosity, as well as self-shielding and self-gravity, in three-dimensional high-resolution numerical simulations. 

The structure of this paper is as follows. We describe the relevant physics equations, the simulation code and initial conditions, and the range of parameters surveyed, in \S~\ref{sec:sims}. Using our suite of simulations and analytic scalings, we then isolate various parameter regimes which give rise to {\em qualitatively} different behaviors in \S~\ref{sec:results:regimes}. We focus on the ``classical cloud destruction'' regime in \S~\ref{sec:regime:destruction:tlife}: there we parameterize the dependence of the cloud lifetime  on the different physical parameters described above, and discuss the effects of different physics. We summarize and conclude in \S~\ref{sec:conclusions}.

\begin{footnotesize}
\ctable[caption={{\normalsize Definitions of variables used in this paper}\label{tbl:variables}},center]{ll}{\tnote[]{}}{
\hline\hline 
$x_{\rm h}$ & value of quantity $x$ in the hot, ambient medium \\ 
$x_{\rm cl}$ & value of quantity $x$ in the cool cloud \\ 
$t_{\rm cool}$ & cooling time $=(3/2)\,k_{B}\,T/n\,\Lambda$ \\ 
$\Lambda$ & cooling function \\ 
$\kappa_{\rm cond}$ & conduction coefficient (see Eq. \ref{eqn:kappa.gizmo})\\
$\nu_{\rm visc}$ & viscosity coefficient (see Eq. \ref{eqn:nu.visc})\\
$\ln{\Lambda_{D}}$ & Coulomb logarithm ($\Lambda_{D}\sim n_{e}\lambda_{D}^{3}$)\\
$n_{e}$ & electron number density \\
$\beta$ & plasma $\beta\equiv P_{\rm therm}/P_{B}$\\
$P_{\rm therm}$ & thermal pressure $=n\,k_{B}\,T$\\
$P_{\rm B}$ & magnetic pressure $=|{\bf B}|^{2}/8\pi$ \\ 
$\chi$ & density contrast $n_{\rm cl}/n_{\rm h}$ ($=T_{\rm h}/T_{\rm cl}$, in equilibrium) \\ 
$c_{s}$ & thermal sound speed \\
$\mathcal{M}_{\rm h}$ & initial Mach number of the hot medium\,$\equiv v_{\rm cl}/c_{s,\,{\rm h}}$ \\
$t_{\rm cc}$ & classical cloud-crushing time $\equiv \chi^{1/2}\,R_{\rm cl}/v_{\rm cl}$ \\
$t_{\rm life,\,pred}$ & predicted cloud lifetime from power-law fit \\
$t_{\rm life,\,sim}$ & simulated cloud lifetime \\
$P_{\rm ram}$ & ram pressure of the ambient medium $=\mu\,m_{\rm p}\,n_{\rm h}\,v_{\rm cl}^{2}$\\
\hline\\
}
\end{footnotesize}

\section{Methods}
\label{sec:sims}
\subsection{Overview \&\ Equations Solved} 

We wish to study the problem of a cloud moving through the ambient CGM. Within the cloud (ignoring, for now, the boundary and shock layer with the hot medium), ideal MHD should be a good approximation but the cooling times are short compared to other macroscopic timescales ($t_{\rm cool} \sim 6\times10^{-5}\,{\rm Myr}$), so we expect clouds to be approximately isothermal at $\sim 10^{4}\,$K (if they are not self-shielding, in which case they might be colder). In the hot medium, on the other hand, radiative cooling is usually negligible over the timescales we consider, as is self-gravity, but the deflection lengths (mean free paths) of the electrons and ions are not negligible. Because the electron and ion gyro-radii are vastly smaller than all other scales in the system, the system can be reasonably described by including appropriate, anisotropic conductive and viscous diffusion coefficients (``Braginskii'' conduction and viscosity; \citealp{1965RvPP....1..205B}), which can provide a reasonable description of the kinetic physics at play (see e.g.,\ discussion in \citealt{Squire2019}).  Indeed, for the regimes considered, transport coefficients perpendicular to the magnetic field are suppressed by factors of $\sim10^{-8}$ compared to the parallel coefficients. Given the large ionization fractions -- $f_{\rm ion}\sim 0.01-1$ inside the cloud, and $f_{\rm ion}\approx 1$ outside it -- we can safely neglect the effect of ambipolar diffusion, the Hall effect, and Ohmic resistivity on the evolution of the magnetic field.

The system of fluid equations we solve is therefore given by:
\begin{align}
\label{eqn:advection}\frac{\partial \rho}{\partial t} + \nabla\cdot\left( \rho\,\mathbf{v} \right) =&\, 0 \\
\label{eqn:momentum}\frac{\partial \mathbf{v}}{\partial t} + \left(\mathbf{v}\cdot \nabla \right)\,\mathbf{v}  =&\, \frac{1}{\rho}
\nabla \cdot {\bf S}
 - \nabla\Phi \\ 
\label{eqn:energy}\frac{\partial e}{\partial t} + \nabla \cdot \left( e\,\mathbf{v} \right) =&\, \nabla \cdot \left( {\bf S}\cdot{\bf v} + {\bf K} \cdot \nabla T \right) - \rho\,{\bf v}\cdot \nabla\Phi - n^{2}\,\Lambda  \\ 
\frac{\partial {\bf B}}{\partial t} =&\, \nabla \times \left( {\bf v} \times {\bf B} \right) \\
\nabla^{2} \Phi =& \,4\pi\,G\,\rho \\ 
{\bf S} \equiv& \, \left( P + \frac{{\bf B}\cdot {\bf B}}{2} \right)\,{\bf I} - {\bf B}\otimes {\bf B} -\boldsymbol{\Pi}  \\ \label{eqn:e.defn}e \equiv& \frac{1}{(\gamma-1)}\,P + \frac{1}{2}\,\rho\,{\bf v}\cdot {\bf v} +\frac{{\bf B}\cdot {\bf B}}{2}
\end{align}
These are the usual continuity, momentum, energy, induction, Poisson (self-gravity) equations, for the gas mass density $\rho$, velocity ${\bf v}$, energy $e$, gravitational potential $\phi$, and magnetic field ${\bf B}$\footnote{To maintain $\nabla \cdot$ {\bf B} = 0, we adopt the divergence cleaning scheme proposed in \citet{Dedner2002} and the constrained gradient scheme in \citet{divergence2016}.}. Here ${\bf S}$ is the stress tensor, with $P = n\,k_{B}\,T$ the usual isotropic (thermal) pressure ($T$ the temperature and $n = \rho/\mu$ the particle number density, with local adiabatic index $\gamma=5/3$).  The conductivity (${\bf K}$) and the viscous part of the stress tensor ($\boldsymbol{\Pi}$) are given by \citet{1953PhRv...89..977S} and \citet{1965RvPP....1..205B} as: 
\begin{align}
{\bf K} &\equiv \kappa_{\rm cond}\,\hat{B}\otimes \hat{B} \\
\kappa_{\rm cond} &= \frac{0.96\,f_{i}\,(k_{B}T)^{5/2}\,k_{B}}{m_{e}^{1/2}e^{4}\ln{\Lambda_{D}}}\,\left({1+4.2\,\ell_{e}/\ell_{T}}\right)^{-1} \label{eqn:kappa.gizmo}\\
\boldsymbol{\Pi} &\equiv 3\,\nu_{\rm visc}\,\left( \hat{B}\otimes\hat{B} - \frac{1}{3}{\bf I}\right)\,\left[ \left( \hat{B}\otimes\hat{B} - \frac{1}{3}{\bf I} \right) : \left( \nabla\otimes {\bf v} \right) \right] \\ 
\label{eqn:nu.visc} \nu_{\rm visc} &= \frac{0.406\,f_{i}\,m_{i}^{1/2}\,(k_{B}T)^{5/2}}{(Z_{i}\,e)^{4}\ln{\Lambda_{D}}}\,\left({1+4.2\,\ell_{i}/\ell_{v}}\right)^{-1} 
\end{align}
where $\otimes$ denotes the outer product; ${\bf I}$ is the identity matrix; ``$:$'' denotes the double-dot-product (${\bf A}:{\bf B} \equiv {\rm Trace}({\bf A}\cdot{\bf B})$); $\ln{\Lambda_{D}}\approx 37.8$ from \citet{1988xrec.book.....S}; $m_{e}$, $e$, $m_{i}$, $Z_{i}\,e=e$ are the electron mass and charge and ion mass and charge; $f_{i}$ the ionized fraction (calculated self-consistently in our cooling routines); $k_{B}$ the Boltzmann constant; $\ell_{e} \approx 0.73\,(k_{B}T)^{2}/(n_{e}\,e^{4}\,\ln{\Lambda_{D}})$ is the electron mean-free path and $\ell_{T} = T/|\nabla T|$ the temperature gradient scale length ($\ell_{i}$ and $\ell_{v} = |{\bf v}|/||\nabla \otimes {\bf v}||$ are the ion mean-free path and velocity gradient scale length). These additional terms  account for saturation of $\kappa$ or $\nu$, although, due to the current uncertainty in the relevant physics, they neglect the effect of plasma ``micro-instabilities,'' which can act to limit the flux further in the high-$\beta$ regime \citep[e.g.,][]{Kunz2014,Komarov2016}. At a sharp discontinuity -- for example, the contact discontinuity at the edge of the cloud --  the form of Eq. \eqref{eqn:kappa.gizmo} ensures the conductive flux takes the  saturated form from \citet{1977ApJ...211..135C}: $q_{\rm sat} \approx 0.4\,(2\,k_{B}\,T/\pi\,m_{e})^{1/2}\,n_{e}\,k_{B}\,T\,\cos{\theta}\,\hat{B}$ (where $\theta$ is the angle between ${\bf B}$ and $\nabla T$). Note, however, that by solving a single set of fluid equations we are assuming that ions and electrons maintain similar temperatures, despite the species having  different conductive heat fluxes. Finally, $\Lambda=\Lambda(T,\,n,\,Z,\,I_{\nu},\,...)$ represents cooling {\em and} heating (so it can have either sign) via additional processes such as radiation, cosmic rays, dust collisions and photo-electric processes, etc (details below). 

\subsection{Simulation Code}
\label{sec:sims:code}

We solve the equations (\ref{eqn:advection})-(\ref{eqn:nu.visc}) in the code {\small GIZMO} \citep{2015MNRAS.450...53H}\footnote{A public version of this code is available at \gizmourl.}, which uses a Lagrangian mesh-free finite-volume Godunov method, in its meshless finite-volume (finite-element) ``MFV'' mode. We have also compared simulations using {\small GIZMO} with its meshless finite-mass, or fixed-grid finite volume solvers, to verify that the choice of hydrodynamic solver in {\small GIZMO} has only small effects on our results. \citet{2015MNRAS.450...53H}, \citet{2016MNRAS.455...51H}, and \citet{divergence2016, 2017MNRAS.466.3387H} present details of these methods and extensive tests of their accuracy and convergence in good agreement with state-of-the-art grid codes (e.g.,\ {\small ATHENA}). In particular the MFV method is manifestly conservative of mass, momentum, and energy, with sharp shock-capturing and accurate treatment of fluid-mixing instabilities (e.g.,\ Kelvin-Helmholtz (KH) and Rayleigh-Taylor (RT) instabilities), and correctly captures MHD phenomena including the magneto-rotational instability (MRI), magnetic jet launching in disks, magnetic fluid-mixing instabilities, and sub-sonic and super-sonic MHD turbulent dynamos. In \citet{2017MNRAS.466.3387H}, we show that the numerical implementation of the anisotropic diffusion operators (${\bf K}$ and $\boldsymbol{\Pi}$) is accurate, able to handle arbitrarily large anisotropies, converges comparably to higher-order fixed-grid codes, and is able to correctly capture complicated non-linear instabilities sourced by anisotropic diffusion such as the magneto-thermal and heat-flux buoyancy instabilities; this has also been tested in fully non-linear simulations of galaxy and star formation \citep{2017MNRAS.471..144S}. {\small GIZMO} also includes full self-gravity ($\phi$) using an improved version of the Tree-PM solver from {\small GADGET-3} \citep{2005MNRAS.364.1105S}, with fully-adaptive and conservative gravitational force softenings (so hydrodynamic and gravitational force resolution is self-consistently matched) following \citet{2007MNRAS.374.1347P}. Finally, {\small GIZMO} includes a detailed, fully-implicit solver for radiative heating and cooling ($\Lambda$). We use the cooling physics from the cosmological FIRE galaxy simulations, with all details given in Appendix B of \citet{2018MNRAS.480..800H}: cooling is tracked self-consistently from $10-10^{10}\,$K, including free-free, photo-ionization/recombination, Compton, photoelectric \&\ dust collisional, cosmic ray, molecular, and metal-line \&\ fine-structure processes (tabulated from {\small CLOUDY}; \citealt{1998PASP..110..761F}) from each of 11 species, accounting for photo-heating by a meta-galactic UV background (using the $z=0$ value from \citealt{2009ApJ...703.1416F}), with self-shielding \citep[as in][]{2013MNRAS.430.2427R} and optically-thick cooling. Additional details are provided in \citet{2018MNRAS.480..800H}; the cooling physics have been used extensively in simulations of star and galaxy formation in the FIRE project. Ionization states are calculated self-consistently accounting for both collisional and photo-ionization. 

\begin{footnotesize}
\ctable[caption={{\normalsize Parameters varied}\label{tbl:parameters}},center]{lll}{\tnote[ ]{The description and parameter space of the main physical parameters varied in this paper.
}}{
\hline\hline 
Name & Description & Values considered \\ 
\hline
$L_{\rm cl}$ & initial cloud diameter (=2$R_{\rm cl}$) & $0.01$, $0.1$, $1$, $10$, $100$, $1000\,{\rm pc}$ \\
$v_{\rm cl}$ & initial cloud velocity & $10$, $100$, $1000\,{\rm km\,s^{-1}}$ \\ 
$T_{\rm h}$ & ambient temperature & $10^{5}$, $10^{6}$, $10^{7}\,$K \\ 
$n_{\rm h}$ & ambient density & $10^{-4}$, $10^{-3}$, $10^{-2}$, $10^{-1}\,{\rm cm^{-3}}$ \\ 
\hline
}
\end{footnotesize}

\subsection{Initial Conditions \&\ ``Default'' Problem Setup}
\label{sec:sims:ics}

Our simulations follow a standard ``cloud crushing'' problem setup, always in three dimensions. For simplicity, a spherical cloud of radius $R_{\rm cl}$ and mean density $n_{\rm cl} \equiv M_{\rm cl} / (4\pi/3\,R_{\rm cl}^{3}\,m_{p}) $ is initialized at an equilibrium temperature $T_{\rm cl} \sim 10^{4}\,$K (with heating and cooling from the meta-galactic UV background), in pressure equilibrium with a homogeneous box filled with gas at electron density $n_{e} = n_{\rm h}$, temperature $T_{\rm h}$, and relative velocity ${\bf v} = v_{\rm cl}\,\hat{y}$ to the cloud (we relax the cloud before turning on velocities to ensure equilibrium temperature and pressure\footnote{We confirm that the cloud expansion during the relaxation process is negligible and dose not affect the subsequent cloud evolution.}). The system is contained in a periodic box with size-length $10\,R_{\rm cl}$ in the $\hat{x}$ and $\hat{z}$ directions and $20\,R_{\rm cl}$ in the $\hat{y}$ direction, with an inflow boundary on the ``upwind'' $\hat{y}$ side such that the upwind portion of the box is always filled with gas at the initial ambient properties (with outflow out of the opposite $\hat{y}$ side). The box moves with the cloud meaning that we can follow the system over long evolution times\footnote{Every time when the cloud material gets too close to the boundary of the box, we shift the entire box to accommodate the cloud again.}, as long as the cloud does not become sufficiently elongated that it exceeds the box size. We have run simulations with box sizes up to $\sim$ 100\,$R_{\rm cl}$ in length to verify that this does not affect our conclusions. One advantage of our Lagrangian code is that it makes no difference (to machine precision) whether we assign the velocity to the cloud or ambient medium. 

In our ``default'' simulations, the box is populated with equal-mass resolution elements with $m_{i}\approx 10^{-6}\,M_{\rm cl}$. Because the method is Lagrangian, our mass resolution is fixed but spatial resolution is automatically adaptive with $\Delta x_{i} \approx 0.01\,R_{\rm cl}\,(n/n_{\rm cl})^{-1/3}\,(m_{i}/10^{-6}\,M_{\rm cl})^{1/3}$. In some of the simulations below we disable self-shielding\footnote{We account for self-shielding following \citet{Faucher15} by locally attenuating the UV background. So to disable self-shielding we simply unattenuate the UV background.} and self-gravity: without self-shielding there is effectively a temperature floor of $\sim 10^{4}\,$K set by the UV background, while with self-shielding gas can cool to $\sim 10\,$K in principle. The default simulations initialize an intentionally weak uniform magnetic field with $\beta \equiv P_{\rm therm}/P_{B} = 10^{6}$, oriented perpendicular to the cloud velocity vector, but we vary this below. A small subset of our simulations consider ``turbulent'' initial conditions, as described below. In Appendix~\ref{sec:convergence}, we show the effects of changing resolution ($m_{i} \sim 10^{-7}-10^{-3}\,M_{\rm cl}$) and verify that the predicted cloud lifetimes are robust to the choice of resolution.

Table \ref{tbl:parameters} lists the key physical parameters that we vary between simulations. We survey a wide range of parameters, including $L_{\rm cl}$ from 0.01 to 1000\,pc, $v_{\rm cl}$ from 10 to 1000\,${\rm km\,s^{-1}}$, $T_{\rm h}$ from $10^{5}$ to $10^{7}$\,K, and $n_{\rm h}$ from $10^{-4}$ to $10^{-1}$\,${\rm cm^{-3}}$.

\subsection{Definition of Cloud ``Destruction'' and ``Lifetime''}
\label{sec:sims:lifetime}

Although it is often obvious ``by-eye'' when a cloud is being ``destroyed'' or ``mixed,'' there is no obvious rigorous definition. Following one common convention in  the literature, we simply define the ``cloud mass'' as the mass above some density threshold relative to the background. Since we consider a range of clouds with different initial density contrasts, we specifically define the mass variable $m_{\rm cl,\,x}$ as the mass in the box with density ${\rm log}\rho > {\rm log} \rho_{\rm h}^{0} + (x/100)\,({\rm log}\rho_{\rm cl}^{0} - {\rm log}\rho_{\rm h}^{0})$, where $\rho_{\rm h}^{0}$ and $\rho_{\rm cl}^{0}$ are the initial ambient and cloud mean densities. So $m_{\rm cl,\,50}$ is the mass above a density threshold equal to $(\rho_{\rm cl}^{0} \rho_{\rm h}^{0})^{1/2}$, i.e., the geometric mean of the initial cloud and ambient medium densities. We have experimented with different values of $x$ from $\sim 5-95$, as well as different functional forms for a density threshold and combined density-temperature thresholds. We find that $m_{\rm cl,\,50}$ defined in this manner gives the most robust estimate of the visually-identified ``cloud'' material, so we will adopt this by default throughout. 

\begin{figure}
\includegraphics[width=0.49\textwidth]{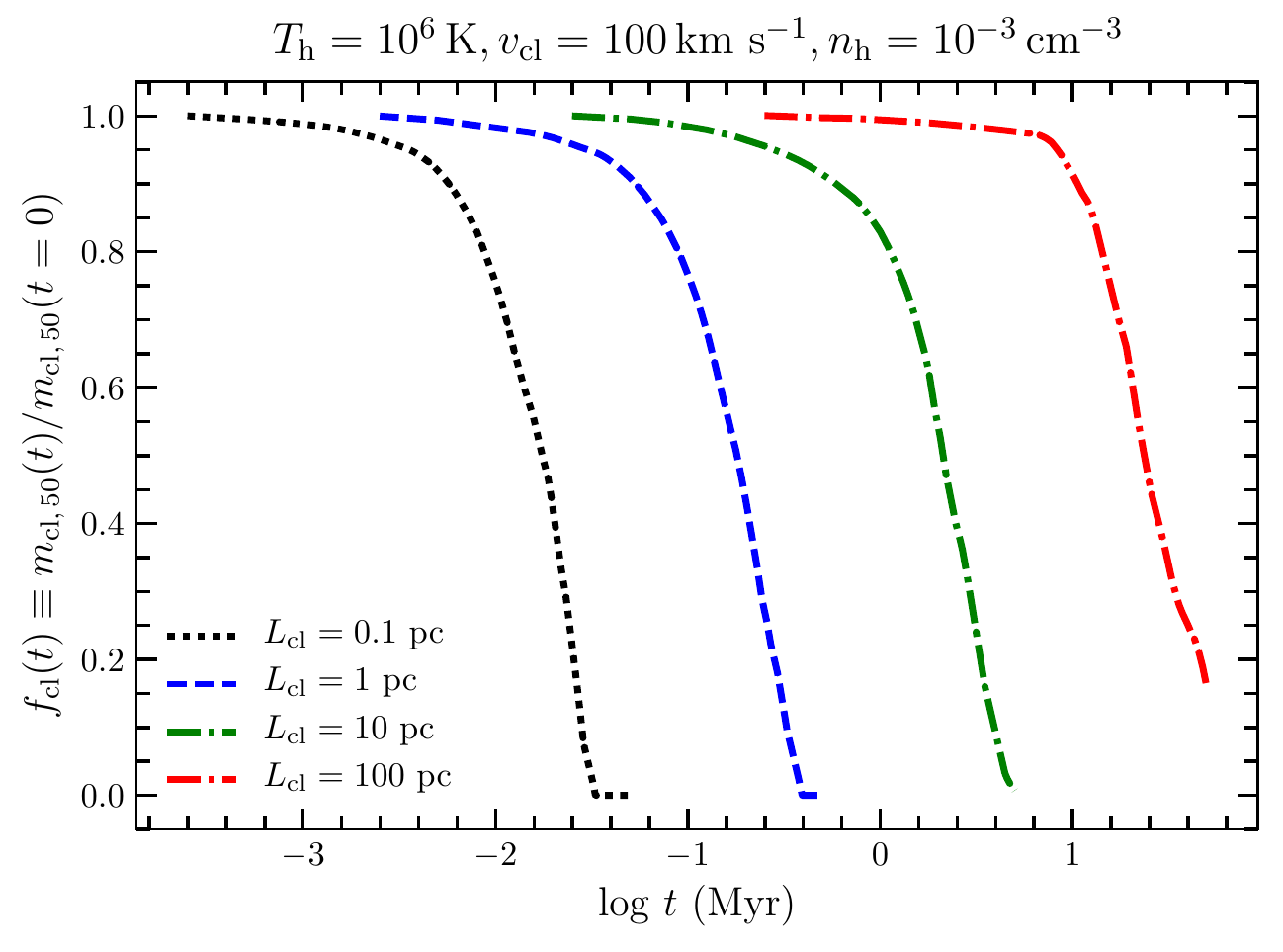} 
    \vspace{-0.25cm}
    \caption{Time evolution of the normalized cloud masses, $f_{\rm cl}(t)$, for four clouds with initial conditions of $T_{\rm h}$ = 10$^6$ K, $v_{\rm cl}$ = 100 ${\rm km\,s^{-1}}$, $n_{\rm h}$ = 10$^{-3}$ ${\rm cm^{-3}}$ and $L_{\rm cl}$ = 0.1-100 pc. Here $f_{\rm cl}(t)$ is defined as $m_{\rm cl,\,50}(t) / m_{\rm cl,\,50}(t=0)$, where $m_{\rm cl,\,50}$ is the cloud mass with density $\rho > (\rho_{\rm cl}^{0} \rho_{\rm h}^{0})^{1/2}$, i.e., the geometric mean of the initial cloud and ambient medium densities. These clouds ``disrupt'' in a well-defined manner in our simulations. We therefore define a cloud ``lifetime'', $t_{\rm life}$, as the time when the cloud mass falls below $10\%$ of its initial value for the first time, i.e., $f_{\rm cl}(t = t_{\rm life}) \leq 0.1$.
    \label{fig:mcloud_definition}}
\end{figure}

Figure \ref{fig:mcloud_definition} shows several examples of the cloud mass estimator, $f_{\rm cl}(t)\equiv m_{\rm cl,\,50}(t) / m_{\rm cl,\,50}(t=0)$ (cloud mass normalized to the initial cloud mass at time $t=0$), as a function  of time. We see in many  of the cases discussed  below that the cloud mass (mass remaining at high densities) declines steadily with  time. In these cases, it is convenient to define a ``lifetime'' $t_{\rm life}$ of the cloud, although this is again somewhat arbitrary. We define this as the time when $f_{\rm cl}(t = t_{\rm life}) \leq 0.1$ for the first time -- i.e.,\ when the cloud mass as defined above falls below $10\%$ of its initial value. We find this is more stable than fitting, e.g.,\ an exponential or power-law decay timescale, because exponential or power-law decay is often not a good approximation to the simulation results. The choice of $\sim 10\%$ of the initial mass is arbitrary, but our results are qualitatively identical for choices in the range $\sim 1-50\%$ (above $\sim 50\%$, we find we often under-estimate the lifetimes of clouds, as  they partially disrupt or evaporate but retain a long-lived ``core,'' and below $\sim 1-2\%$, resolution concerns begin to dominate). 

Not all clouds decay in mass: as we will show below, some grow. For these, we can define a growth timescale as the approximate $e$-folding time. 

\section{Different Regimes of Dominant Physics}
\label{sec:results:regimes}

Guided by our simulation parameter survey, plus some basic analytic considerations, we now define different regimes of cloud behavior in the CGM and the most relevant physics in each.

\subsection{The Smallest Clouds: Where Conduction Breaks Down}
\label{sec:results:regimes:tiny}

The thermal conductivity of the hot medium is defined by the transport of hot electrons, with $\kappa/k_{B}\,n_{\rm h} \sim \lambda_{e,\,{\rm h}}\,c_{s,\,e,\,{\rm h}}$ where \begin{equation}\lambda_{e,\,{\rm h}} \equiv 3\,m_{e}^{1/2}\,(k_{B}\,T_{e})^{3/2}\,c_{s,\,e,\,{\rm h}} / (4\sqrt{2\pi}\,n_{i}\,e^{4}\,\ln{\Lambda_{D}}) \approx 0.1\,{\rm pc}\,\frac{T_{\rm 6}^2}{n_{\rm h,\,0.01}}\end{equation} 
(using $\ln{\Lambda_{D}}\approx 26$ for $T_{\rm h} \sim 10^5 - 10^6\,\rm K$) is the electron Coulomb deflection length (along the magnetic field) and $c_{s,\,e,\,{\rm h}}$ is the electron isothermal sound speed ($\equiv \sqrt{k_{B}\,T_{\rm h}/m_{e}}$) defined in the hot medium. When the hot electrons encounter a cold cloud, they are able to penetrate to a skin depth $\lambda_{\rm skin} = \lambda_{e,\,{\rm h}}\,(n_{\rm h}/n_{\rm cl}) = \lambda_{e,\,{\rm h}}\,(T_{\rm cl}/T_{\rm h})$. If $\lambda_{\rm skin} \gtrsim R_{\rm cl}$, then our description of heat transport (conduction) via Eq. (\ref{eqn:energy}) breaks down (regardless of the accounting for saturated v.s.\ unsaturated conduction). Using the values above, this occurs when 
\begin{align}
\label{eqn:NH.small}
N_{\rm H} \lesssim N_{\rm H}^{\rm mfp} &\sim 10^{16}\,{\rm cm^{-2}}\,T_{6}^{2}
\end{align}
where $T_{6} \equiv T_{\rm h}/10^{6}\,$K, and $N_{\rm H} \equiv R_{\rm cl}\,\langle n_{\rm cl} \rangle$ is the column density through the cloud\footnote{It is sometimes stated that the ``fluid approximation'' breaks down on scales small compared to $\lambda_{\rm skin}$ or even the (much larger) $\lambda_{e,\,{\rm h}}$, but this is not necessarily correct. So long as the gyro radii of the particles remain small compared to the relevant scales, equations with a similar form to the fluid MHD equations (the ``kinetic MHD'' equations of \citealt{Kulsrud1983}) remain valid.  However, our descriptions of parallel heat and momentum transport clearly become problematic below $\lambda_{\rm skin}$, as does the assumption that the electrons and ions remain at the same temperature.}.

We therefore intentionally avoid simulating systems below this scale. However, we can estimate what will occur. In this limit, the free $e^{-}$ in the hot medium  effectively do not ``see'' the cloud: the cloud will effectively be immersed in a sea of hot $e^{-}$ with number density equal to the ambient hot $e^{-}$ density,  which contribute a uniform volumetric Coulomb heating rate. If the cloud is ionized, this is just $\dot{e} = 0.34\,n_{e,\,{\rm h}}\,(c_{s,\,e,\,{\rm h}}/\lambda_{\rm skin})\,k_{B}\,T_{\rm h}$ \citep{2016ApJ...822...31B}, and if $T_{6}^{3/2} \Lambda_{\rm cl,-23}\lesssim 0.14$, then the volumetric heating rate from hot $e^{-}$ is larger than the cooling rate of gas in the cloud, and they should evaporate on a timescale short compared to their sound-crossing times.  This process is analysed in detail in \citet{Balbus1982}.

\subsection{Self-gravity \&\ Self-Shielding}
\label{sec:results:regimes:grav.shield}

\begin{figure}
\includegraphics[width=0.487\textwidth]{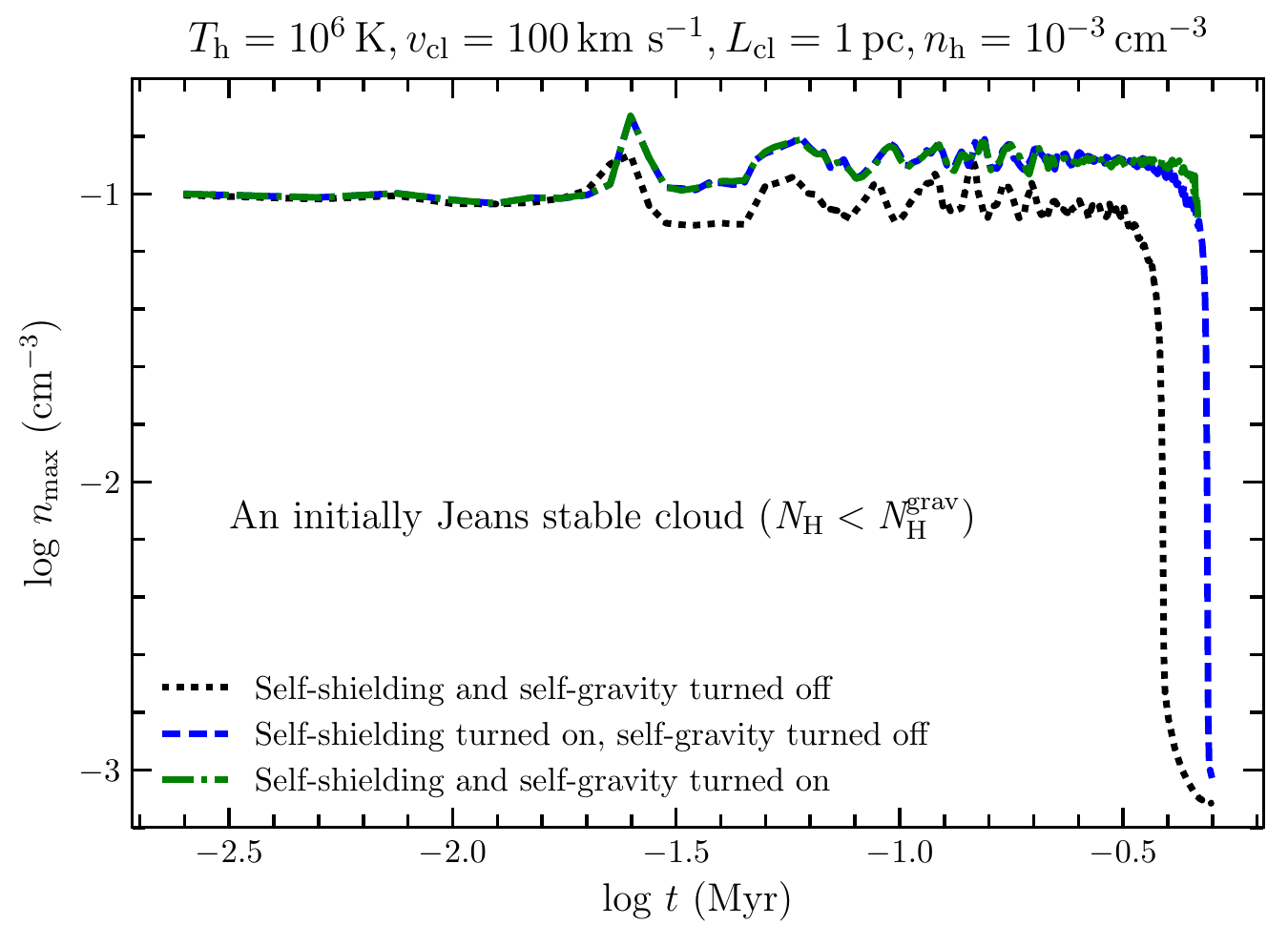}
\includegraphics[width=0.49\textwidth]{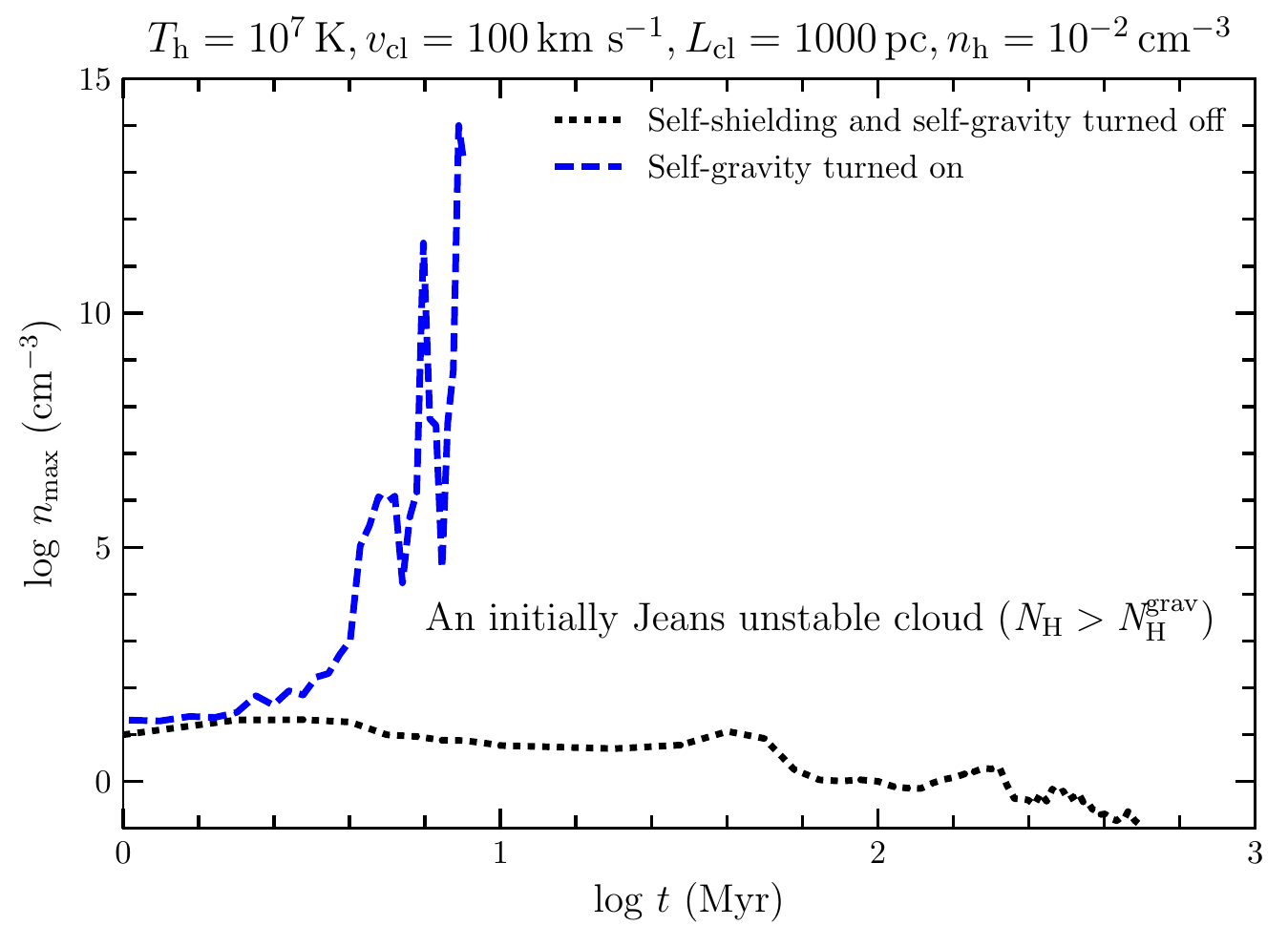}
    \vspace{-0.25cm}
    \caption{Time evolution of the maximum density ($n_{\rm max}$) in a cloud for two representative cases. \emph{Upper:} If $N_{\rm H} \lesssim N_{\rm H}^{\rm grav}$ (Eq. \ref{eqn:NH.jeans}), i.e., the cloud is initially Jeans-stable, then turning on or off self-gravity or self-shielding makes little difference. \emph{Lower:} If $N_{\rm H} \gtrsim N_{\rm H}^{\rm grav}$ (the cloud is initially Jeans-unstable), turning on self-gravity leads to cloud collapse ($n_{\rm max}$ runs away) in a free-fall time, as expected.
    \label{fig:SSSG}}
\end{figure}

At the other extreme, consider very large clouds. If a cloud is {\em initially} self-gravitating/Jeans-unstable, i.e.,\ has $\lambda_{\rm J} \equiv c_{s,\,{\rm cl}}/\sqrt{G\,\rho_{\rm cl}} \ll R_{\rm cl}$, or $R_{\rm cl} \gtrsim 1\,{\rm kpc}\,(n_{\rm h,\,0.01}\,T_{6})^{-1/2}$, or 
\begin{align}
\label{eqn:NH.jeans} N_{\rm H} \gtrsim N_{\rm H}^{\rm grav} &\sim 0.5 \times 10^{22}\,{\rm cm^{-2}}\,(n_{\rm h,\,0.01}\,T_{6})^{1/2} \\
& \sim 10^{22}\,{\rm cm^{-2}}\,P_{-12}^{1/2} \nonumber
\end{align}
where $P_{-12} \equiv P_{\rm h}/10^{-12}\,{\rm erg\,cm^{-3}}$, then (a) the gravitational force per unit area is larger than the external (confining/stripping) pressure, and (b) its collapse/free-fall time is shorter than its sound-crossing time, itself shorter than the cloud  destruction time (in the absence of gravity). Figure \ref{fig:SSSG} shows that in our simulations with self-gravity on, we confirm that clouds which are initially Jeans-unstable ($N_{\rm H} > N_{\rm H}^{\rm grav}$; Eq.~\ref{eqn:NH.jeans}) indeed fragment/collapse rapidly\footnote{Since we do not include star formation, we eventually stop the simulations when most of  the  gas in the initial cloud has collapsed to densities $>10^{5}$ times larger than its initial mean density.}, while clouds which are initially Jeans-stable ($N_{\rm H} < N_{\rm H}^{\rm grav}$) behave essentially identically whether or not self-gravity is included. Thus, self-gravity is very much a ``threshold'' effect: it dominates in Jeans-unstable clouds, and is irrelevant in Jeans-stable clouds (at least on the spatial/time scales we simulate). There is only a very narrow, fine-tuned,  and dynamically unstable parameter space where clouds are ``just barely'' Jeans-stable initially and can have sub-regions ``pushed into'' Jeans instability by their interactions  with the ambient medium (we find just one such example in our entire parameter survey, with initial $N_{\rm H} \sim 0.8\,N_{\rm H}^{\rm grav}$)\footnote{This is expected: 1D compression (e.g.,\ the initial ``pancaking'' of the cloud as it shocks) does not strongly enhance Jeans instability. Consider an initially Jeans-stable, isothermal cloud with (pre-shock) Jeans length  $\lambda_{\rm J}^{0}>R_{0}$ (radius $R=R_{0}$), compressed or ``pancaked'' to width $H \ll R_{0}$ along the short axis (retaining $R=R_{0}$ along the long axis). Fragmentation along the short axis requires a Jeans-like criterion $\lambda_{\rm J}^{\rm new} < H$, but $\lambda_{\rm J}^{\rm new} = c_{s}/\sqrt{G\,\rho^{\rm new}} \sim H\,(\lambda_{\rm J}^{0}/R_{0})\,(R_{0}/H)^{1/2} \gg H$. Along the long-axis, fragmentation must be treated two-dimensionally, and requires $\lambda_{\rm J}^{2D} < R_{0}$ where $\lambda_{\rm J}^{2D} \equiv c_{s}^{2}/(\pi\,G\,\Sigma_{\rm cloud}) \sim R_{0}\,(\lambda_{\rm  J}^{0}/R_{0})^{2} \gg R_{0}$. So an initially Jeans-stable cloud remains stable.}. This should not be surprising: the same behavior has been repeatedly demonstrated for clouds in the ISM \citep[see e.g.,][]{1976ApJ...206..753M, 1976ApJ...207..141M, Li2014, Federrath15, Federrath19}. 

Likewise, if the cloud can {\em initially} self-shield to molecular or fine-structure metal-line cooling to temperatures $T \sim 10-100\,{\rm K} \ll 10^{4}\,$K, it will cool to those temperatures very quickly, which will remove its internal pressure support and render it immediately Jeans-unstable (even more so, given the rapid compression by the ambient medium  which would follow). This is well-studied in the ISM context and requires a surface density $\gtrsim 10\,M_{\odot}\,{\rm pc^{-2}}\,(Z_{\odot}/Z)$ \citep[see][for extended discussion]{2008ApJ...680.1083R, 2011ApJ...729...36K}, or a column density
\begin{align}
\label{eqn:NH.shield} N_{\rm H} \gtrsim N_{\rm H}^{\rm shield} &\sim 1.5 \times 10^{22}\,{\rm cm^{-2}}\,Z_{0.1}^{-1}
\end{align}
where $Z_{0.1} \equiv Z / 0.1\,Z_{\odot}$. Like  with self-gravity, we find this is a sharp ``threshold'' effect, not surprising since the self-shielding attenuation ($\propto e^{-\tau}$) is  an extremely strong function of the $N_{\rm H}$, which can vary by orders of magnitude. Usually, self-shielded clouds ($N_{\rm H}>N_{\rm H}^{\rm shield}$; Eq.~\ref{eqn:NH.shield}) are already self-gravitating, but it is largely irrelevant which occurs ``first.'' A self-shielded (but initially Jeans-stable) cloud rapidly becomes Jeans-unstable, while a Jeans-unstable (but non-shielded) cloud collapses isothermally (at $\sim 10^{4}\,$K) until it becomes  self-shielded, then collapses more rapidly (see \citealt{2008ApJ...680.1083R, 2018MNRAS.478.3653O}). Because the criterion here is a simple column-density threshold, it is also obvious that 1D compression of the cloud does not strongly alter its self-shielding. For the sake of completeness and testing our theory of cloud destruction, we have re-run all our simulations {\em without} self-gravity and self-shielding, so we can see whether and ``how fast'' they would be destroyed in the  absence of these physics in our analysis below, but we stress that this is purely a counter-factual exercise.

\subsection{Rapid Cooling of the Hot Medium: Failure of Pressure Confinement}
\label{sec:results:regimes:hot.cooling}
If the hot gas cools faster than the time it takes to cross/envelop the cloud, it cannot maintain meaningful pressure confinement. Even if we add some global (spatially-uniform) heating rate per unit volume or heat conduction in the hot medium, such that the ambient gas {\em equilibrium} temperature remains fixed at the ``target'' temperature, in this limit the hot gas is still thermally unstable and it cannot respond to perturbations of the cloud shape or expansion of the cloud, so the cloud will behave as if it is in an essentially pressure-free medium. This occurs when $t_{\rm cool,\,h} \lesssim t_{\rm cross} \sim R_{\rm cl}/v_{\rm cl}$ (or $R_{\rm cl}/c_{s, \,\rm cl}$ if $v_{\rm cl} \lesssim c_{s,\,\rm cl}$), giving:
\begin{align}
\label{eqn:NH.cool} N_{\rm H} \gtrsim N_{\rm H}^{\rm confine} &\sim 0.5\times 10^{22}\,{\rm cm^{-2}}\,T_{6}^{2}\,v_{100}\,\Lambda_{\rm h,\,-23}^{-1} 
\end{align}
where $v_{100} \equiv v_{\rm cl}/100\,{\rm km\,s^{-1}}$ and $\Lambda_{\rm h,\, x} \equiv \Lambda(n_{\rm h},\,T_{\rm h},\,Z_{\rm h}) / 10^{\rm x}\,{\rm erg\,cm^{3}}$. For $T_{\rm h} \gtrsim 10^{6}\,$K, this requires larger column densities than would already be self-gravitating or self-shielding, so this parameter regime becomes irrelevant. However, when the hot medium is cooler than $\sim 10^{6}\,$K, cooling becomes much more efficient, and the required $N_{\rm H}$ for this regime drops rapidly (to $\gtrsim 10^{18}\,{\rm cm^{-2}}$ at $T_{\rm h} \sim 10^{5}\,$K). In the CGM, this naturally coincides with the virial temperatures below which ``hot halos'' that can maintain a stable virial shock and quasi-hydrostatic pressure-supported gas halo cease to exist. 

\begin{figure}
\includegraphics[width=0.49\textwidth]{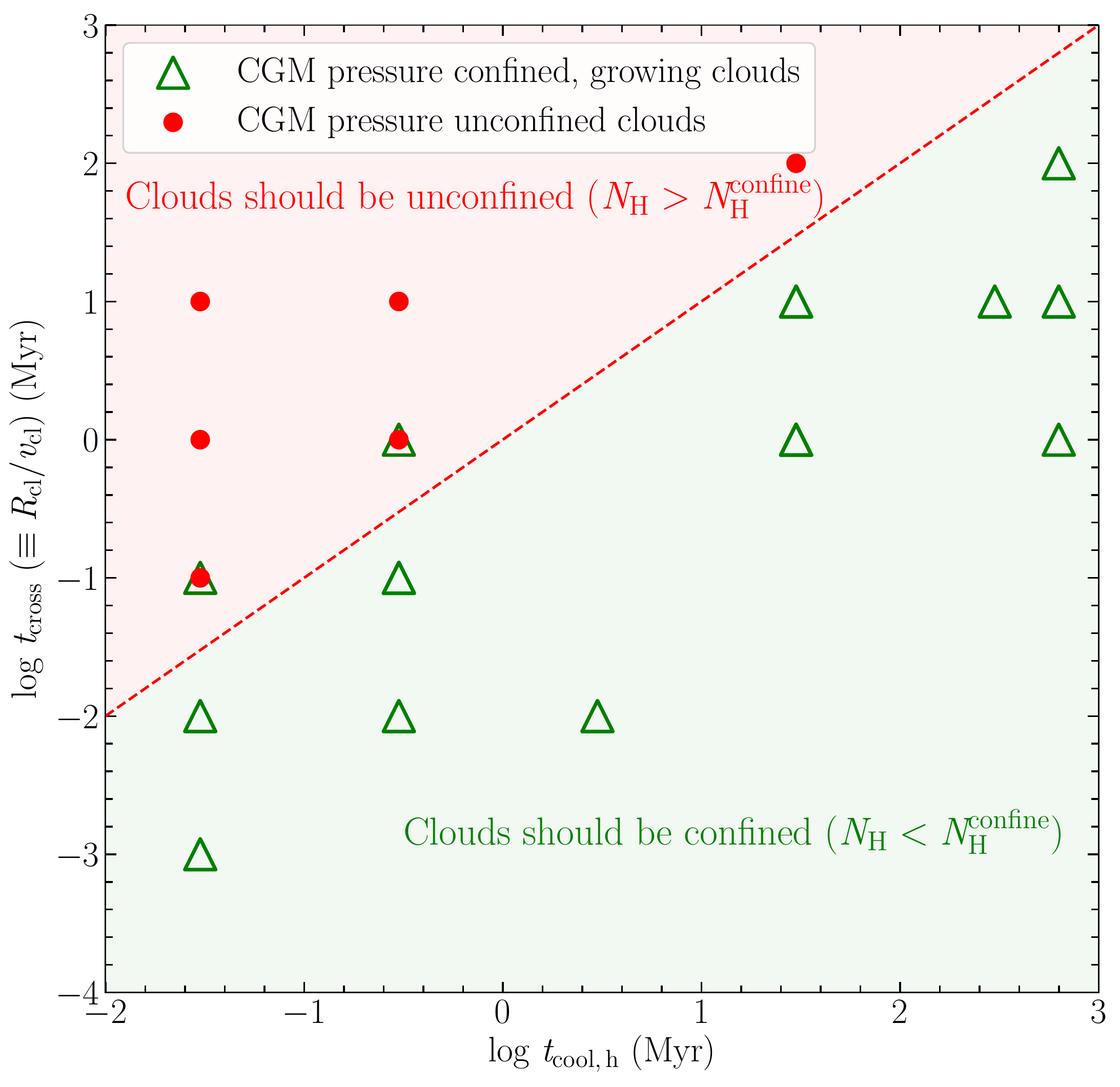}
\includegraphics[width=0.49\textwidth]{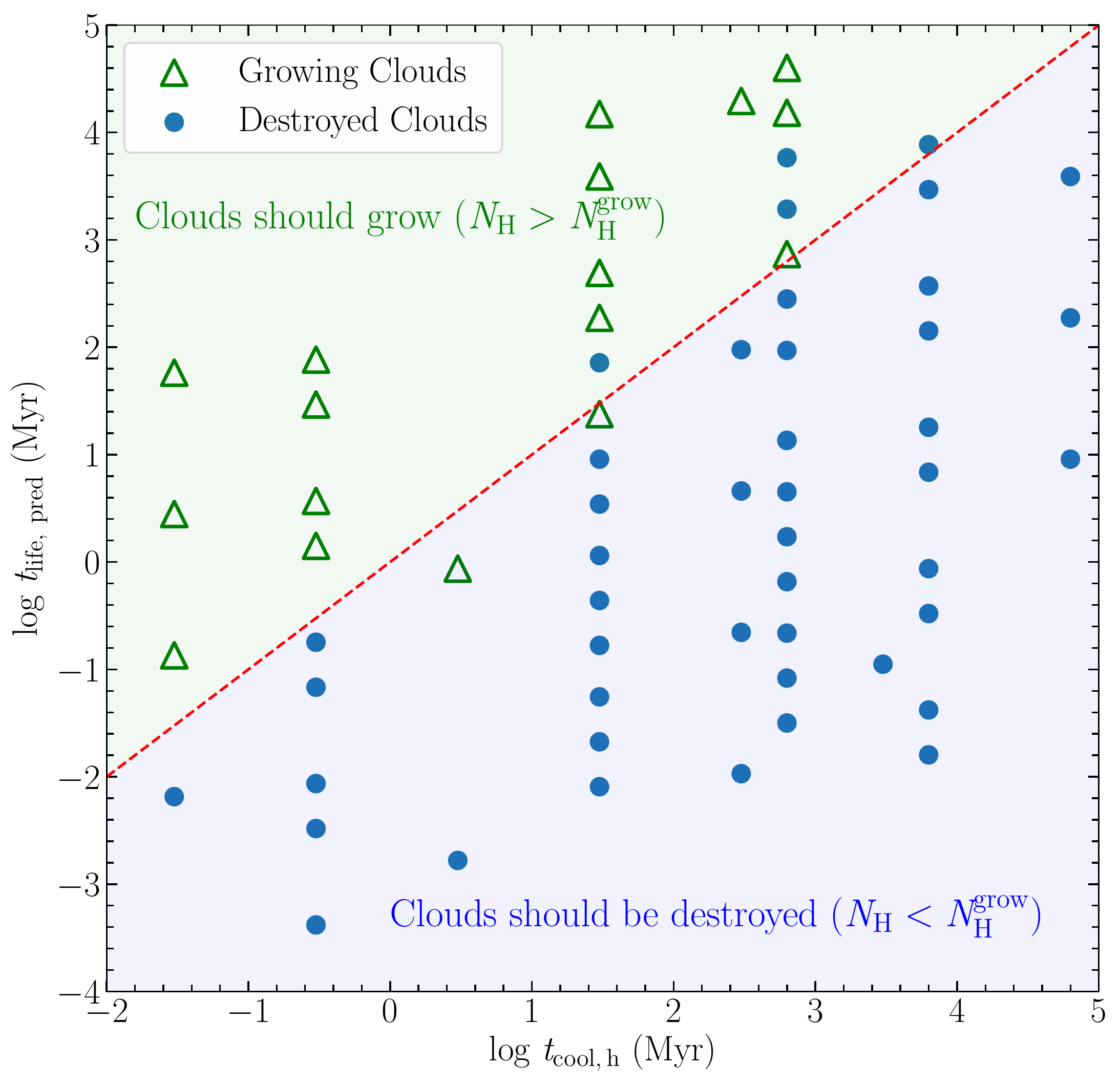}
    \vspace{-0.25cm}
    \caption{Simulation tests of the criteria for separating different cloud behaviors discussed in \S~\ref{sec:results:regimes:hot.cooling} and \S~\ref{sec:results:regimes:growth}. 
    {\em Upper}: Cooling time of ambient hot gas ($t_{\rm cool,\,h}$) v.s.\ crossing time of that gas over the cloud ($t_{\rm cross}$). When cooling is faster than cloud velocity/sound crossing times, the clouds cannot be meaningfully pressure-confined and simply expand (neglecting self-gravity). The green triangles denote simulations used to check this directly, which confirm the validity of the simple analytic criteria for this behavior in Eq.~\eqref{eqn:NH.cool}. 
    {\em Lower:} Same, but comparing $t_{\rm cool,\,h}$ to the cloud ``destruction time'' in the limit where cooling is {\em not} important ($t_{\rm life,\,pred}$, given in \S~\ref{sec:regime:destruction:tlife}, Eq.~\ref{eqn:scaling.relation}). When cooling of the hot gas in the cloud front is faster than cloud disruption, the cloud accretes and grows: simulations confirm the simple analytic criterion derived in Eq.~\eqref{eqn:NH.grow}.
    \label{fig:criteria}}
\end{figure}

\begin{figure}
\includegraphics[width=0.5\textwidth]{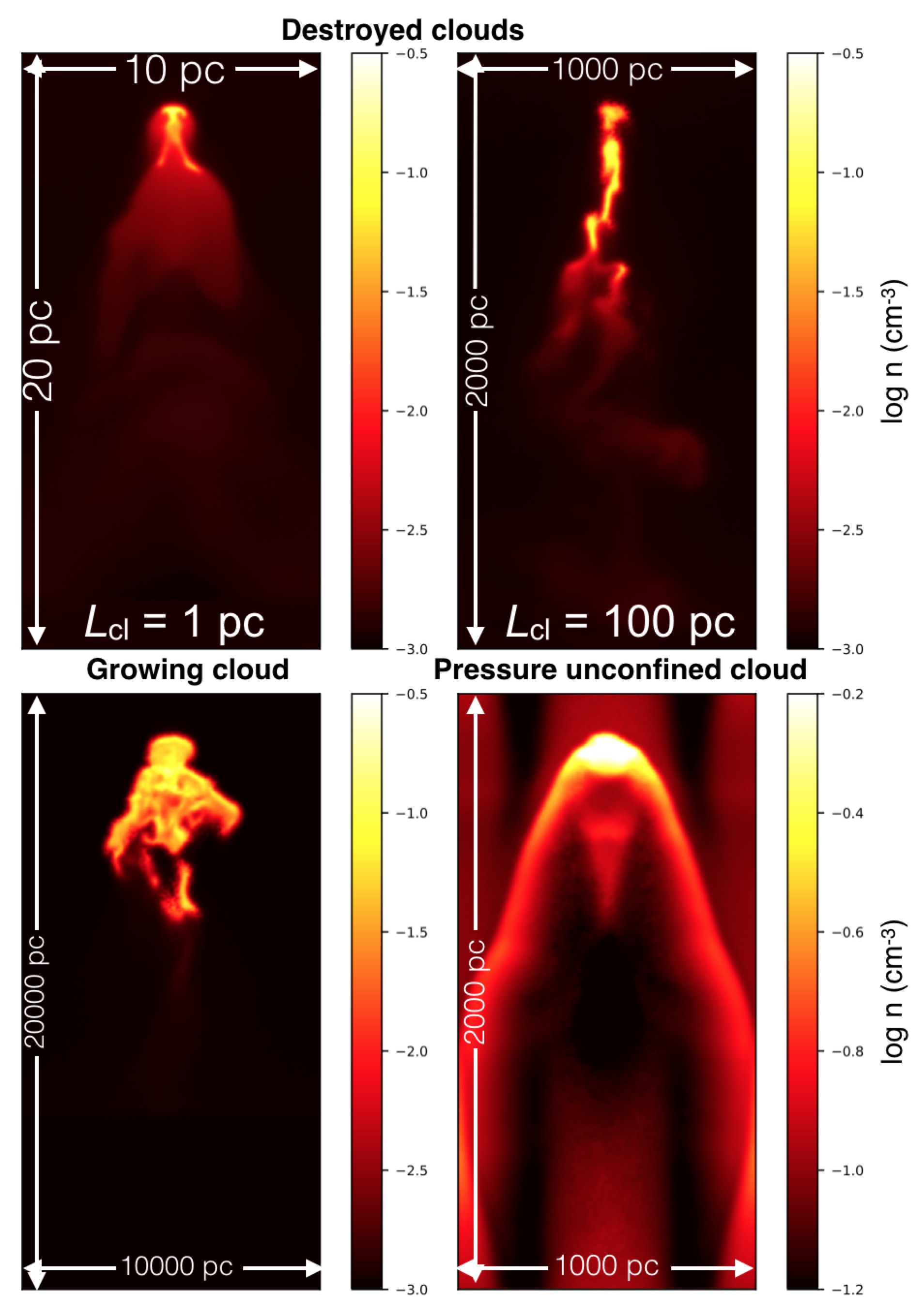}
    \vspace{-0.25cm}
    \caption{{\em Upper}: Sliced density maps of two clouds in the ``classical cloud destruction'' regime with initial conditions of $T_{\rm h}$ = 10$^6$\,K, $v_{\rm cl}$ = 100\,${\rm km\,s^{-1}}$, $n_{\rm h}$ = 10$^{-3}\,{\rm cm^{-3}}$, $L_{\rm cl}$ = 1\, and 100\,pc, respectively. {\em Lower left}: Sliced density map of a ``growing'' cloud ($N_{\rm H} \gtrsim N_{\rm H}^{\rm grow}$, with $T_{\rm h}$ = 10$^6$\,K, $v_{\rm cl}$ = 100\,${\rm km\,s^{-1}}$, $n_{\rm h}$ = 10$^{-3}\,{\rm cm^{-3}}$, $L_{\rm cl}$ = 1000\,pc). {\em Lower right:} Sliced density map of a ``pressure unconfined'' cloud ($N_{\rm H} \gtrsim N_{\rm H}^{\rm confine}$, with $T_{\rm h}$ = 10$^5$\,K, $v_{\rm cl}$ = 100\,${\rm km\,s^{-1}}$, $n_{\rm h}$ = 10$^{-1}\,{\rm cm^{-3}}$, $L_{\rm cl}$ = 100\,pc). \label{fig:CGMcool_snapshots}}
\end{figure}

In Figure \ref{fig:criteria} and \ref{fig:CGMcool_snapshots}, we confirm in our simulations that clouds with $N_{\rm H} \gtrsim N_{\rm H}^{\rm confine}$ (Eq.~\ref{eqn:NH.cool}) indeed behave as if there is negligible confining pressure. As shown in the lower right panel of Figure \ref{fig:CGMcool_snapshots}, they expand into the ambient, low-pressure medium, which does cause the cloud density to decrease, but ambient gas cooling/accretion also causes the cloud mass to grow, so this is clearly distinct from classical cloud ``destruction''. If Eq. \eqref{eqn:NH.cool} is satisfied, the failure of pressure confinement occurs with or without the addition of an artificial spatially-uniform heating rate $Q$ (such that the heating+cooling rate per unit volume is $\dot{e} = Q - n^{2}\,\Lambda$), with $Q$ chosen so the hot gas evolved in isolation (no cold cloud) remains exactly at its initial temperature. While not surprising, this is important for application of our conclusions in the CGM, especially around  dwarf galaxies, which  are in the ``cold  mode'' of accretion without ``hot halos'' \citep{2009ApJ...700L...1K}. In that regime, cold clouds from e.g.,\ galactic winds may well have $N_{\rm H} \gtrsim N_{\rm H}^{\rm confine}$, and thus could behave as if they are expanding into vacuum.

\subsection{Clouds Grow: Accreting Ambient Hot Gas}
\label{sec:results:regimes:growth}

As discussed in recent work by e.g.,\ \citet{Gronke18, Gronke19}, if clouds avoid destruction for a time longer than the cooling time of swept-up material, the front of the hot material entrained by the cloud (and mixing with the denser, cooler, cloud material) cools rapidly and effectively gets ``accreted'' onto the cloud. We can crudely estimate when this occurs by comparing our estimated cloud destruction time via ``shredding'' (in the absence of cooling), $t_{\rm life,\,pred} \sim 10\,t_{\rm cc}\,\tilde{f}$ (defined in \S~\ref{sec:regime:destruction:tlife} below) to the cooling time of the hot medium, $t_{\rm cool,\,h}$.  This gives: 
\begin{align}
\label{eqn:NH.grow}N_{\rm H} \gtrsim N_{\rm H}^{\rm grow} &\sim 2\times10^{20}\,{\rm cm^{-2}}\,T_{6}^{3/2}\,v_{100}\,\tilde{f}^{-1}\,\Lambda_{\rm{h},\,-23}^{-1}
\end{align}
(The material in the front has been heated modestly by compression and/or shocks, but also increased in density, and rapid conduction suppresses temperature variations; thus for the conditions  simulated here the cooling time of the front material is order-unity similar to the cooling time in the ambient gas).
For the range of parameters of interest in the CGM, this {\em almost always} occurs at lower $N_{\rm H}$ compared to the ``failure of pressure confinement'' above. So if a cloud ``begins'' life in-between ($N_{\rm H}^{\rm grow} \lesssim N_{\rm H} \lesssim N_{\rm H}^{\rm confine}$), it will grow until it reaches that larger $N_{\rm H}$ threshold, at which point it will continue to ``sweep up'' any gas in its path, but also expand in the ``backward'' direction as the gas cools around it. Note that, however, if the cloud increases its $N_{\rm H}$ (mass) by an order-unity factor, momentum conservation requires it decelerate by a similar factor. So the cloud will slow down and stop, which in turn decreases $v_{100}$, making it even more above-threshold to survive. So we end up with essentially static, long-lived clouds in this limit.

Note that \citet{Gronke19} derive a criterion for ``cloud growth'' that is slightly different from ours. They start from the same principle, comparing cloud lifetimes and cooling time in the mixing layer/front, but assume the cloud lifetime is  $t_{\rm cc}$ and the cooling time of the ambient hot gas is $t_{\rm cool,\,h}/\chi$ (this arises from assuming the ``near-cloud'' hot gas has geometric-mean temperature and density between cloud and ambient medium, and neglecting the dependence of $\Lambda$ on $T$). Accounting for both efficient conduction and rapid ``sweeping'' of the hot gas past the cloud, we find that simply using $t_{\rm cool,\,h}$ for the ambient gas, together with our more accurate cloud lifetime estimates, provides a more accurate and robust criterion for distinguishing between ``growing'' and ``destroyed'' cloud cases. This is especially true at high ambient temperatures ($T_{\rm h}\gtrsim 10^{6}\,$K), as can be seen in the lower panel of Figure \ref{fig:criteria}. One possible explanation is that efficient conduction heats up the gas in the front and makes it difficult for a mixing layer at intermediate temperature to exist. This effect is shown in the density maps we present in \S\ref{sec:regime:destruction:conduction}, where the cloud with conduction has sharper edges, indicating a sharper density and temperature contrast. Understanding the cause of this discrepancy in more detail will be left to future work.

\begin{figure}
\includegraphics[width=0.49\textwidth]{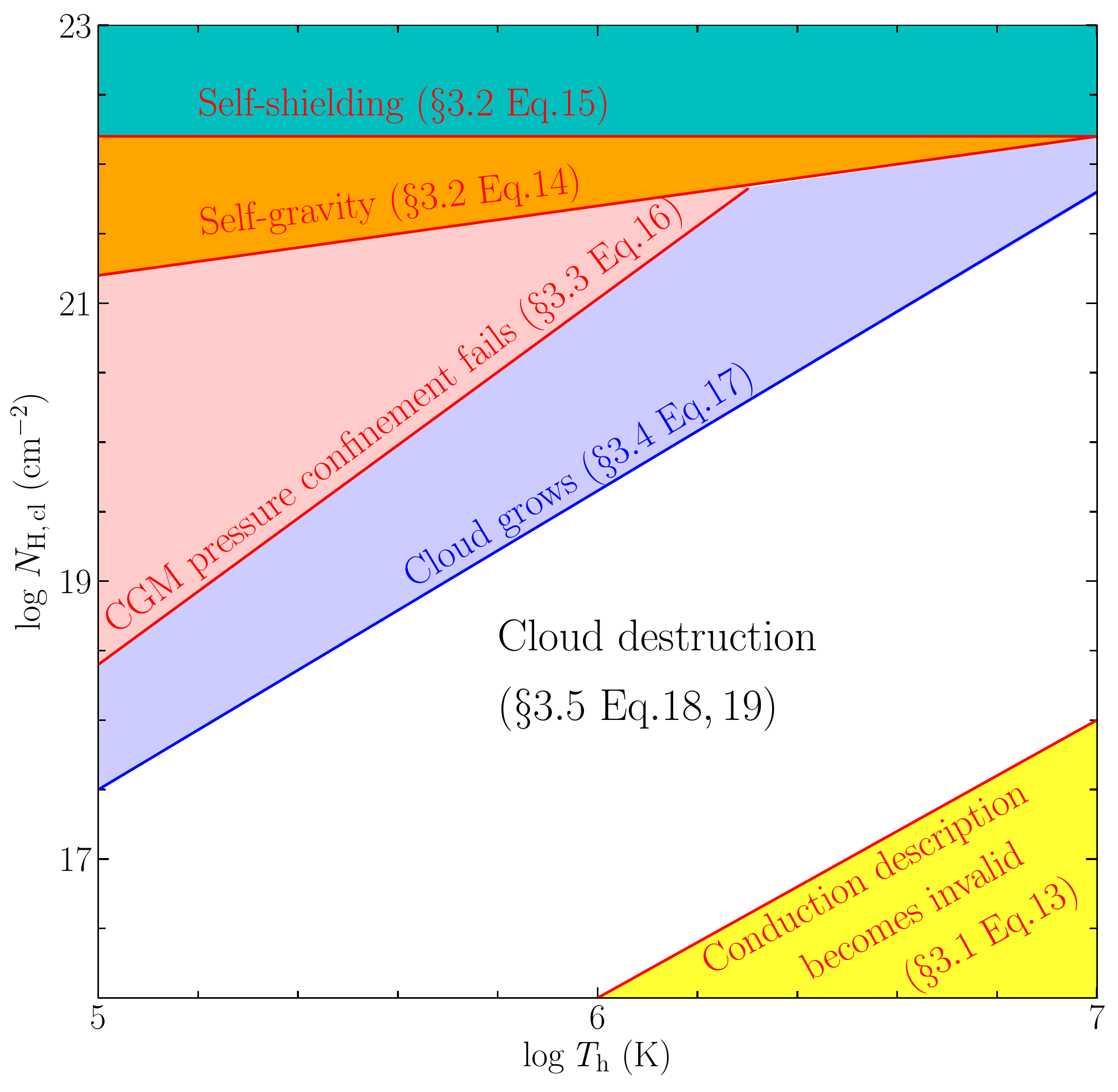}
    \vspace{-0.25cm}
    \caption{The cloud column density ($\rm \emph{N}_{H,\,cl}$) v.s. the temperature of the ambient medium ($T_{\rm h}$). Different regimes of dominant physics are shown: (1) The ``conduction description fails'' regime (\S\ref{sec:results:regimes:tiny}, Eq. \ref{eqn:NH.small}, shown in yellow); (2) The ``self-shielding and self-gravity dominate'' regime (\S\ref{sec:results:regimes:grav.shield}, Eq. \ref{eqn:NH.jeans}, \ref{eqn:NH.shield}, shown in green and orange); (3) The ``CGM pressure confinement fails'' regime  (\S\ref{sec:results:regimes:hot.cooling}, Eq. \ref{eqn:NH.cool}, shown in pink); (4) The ``cloud grows'' regime (\S\ref{sec:results:regimes:growth}, Eq. \ref{eqn:NH.grow}, shown in blue); (5) The ``classical cloud destruction'' regime (\S\ref{sec:regime:destruction:tlife}, Eq. \ref{eqn:NH.destruct}, shown in white). Typical values of certain parameters have been adopted ($v_{\rm cl}$ = 100\,${\rm km\,s^{-1}}$, $n_{\rm h}$ = 10$^{-2}\,{\rm cm^{-3}}$, $\tilde{f}$ = 1).
    \label{fig:regime}}
\end{figure}

\subsection{In-Between: Classical Cloud ``Destruction'' (Shredding)}\label{sec:regime:destruction:tlife}

If we exclude all of the regimes above, i.e.,\ consider only clouds with 
\begin{align}
\label{eqn:NH.destruct}N_{\rm H}^{\rm mfp} \ll N_{\rm H} \ll {\rm min}\left\{N_{\rm H}^{\rm grow},\,N_{\rm H}^{\rm confine},\,N_{\rm H}^{\rm shield},\,N_{\rm H}^{\rm grav}\right\}
\end{align}
then we find that all the clouds we simulate are eventually destroyed/dissolved. The boundaries of this parameter space (where clouds are destroyed) are illustrated in a simple ``contour'' form in Figure \ref{fig:regime}. We find that all clouds in this regime can be at least order-of-magnitude described by traditional cloud-crushing arguments \citep{Klein1994}. This conclusion holds regardless of the specific physics included in a given simulation (e.g., conduction, or self gravity), with the classical cloud-crushing estimate $t_{\rm cc} \sim \chi^{1/2}\,R_{\rm cl}/v_{\rm cl}$ providing a reasonable qualitative starting point to understand the actual cloud destruction times in the simulations. The majority of this section is dedicated to explaining why this is the case.

\begin{figure*}
\includegraphics[width=0.49\textwidth]{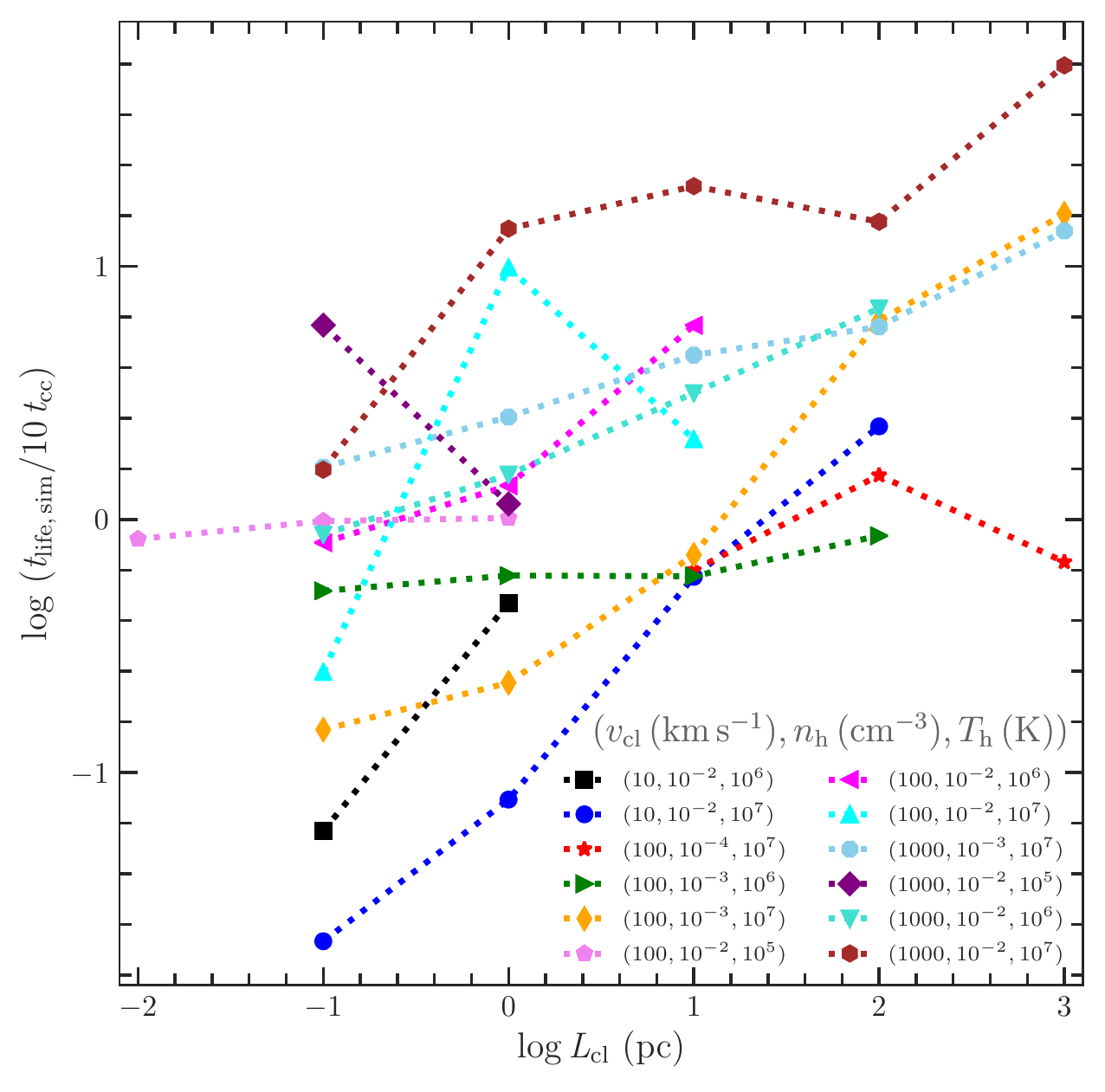}
\includegraphics[width=0.49\textwidth]{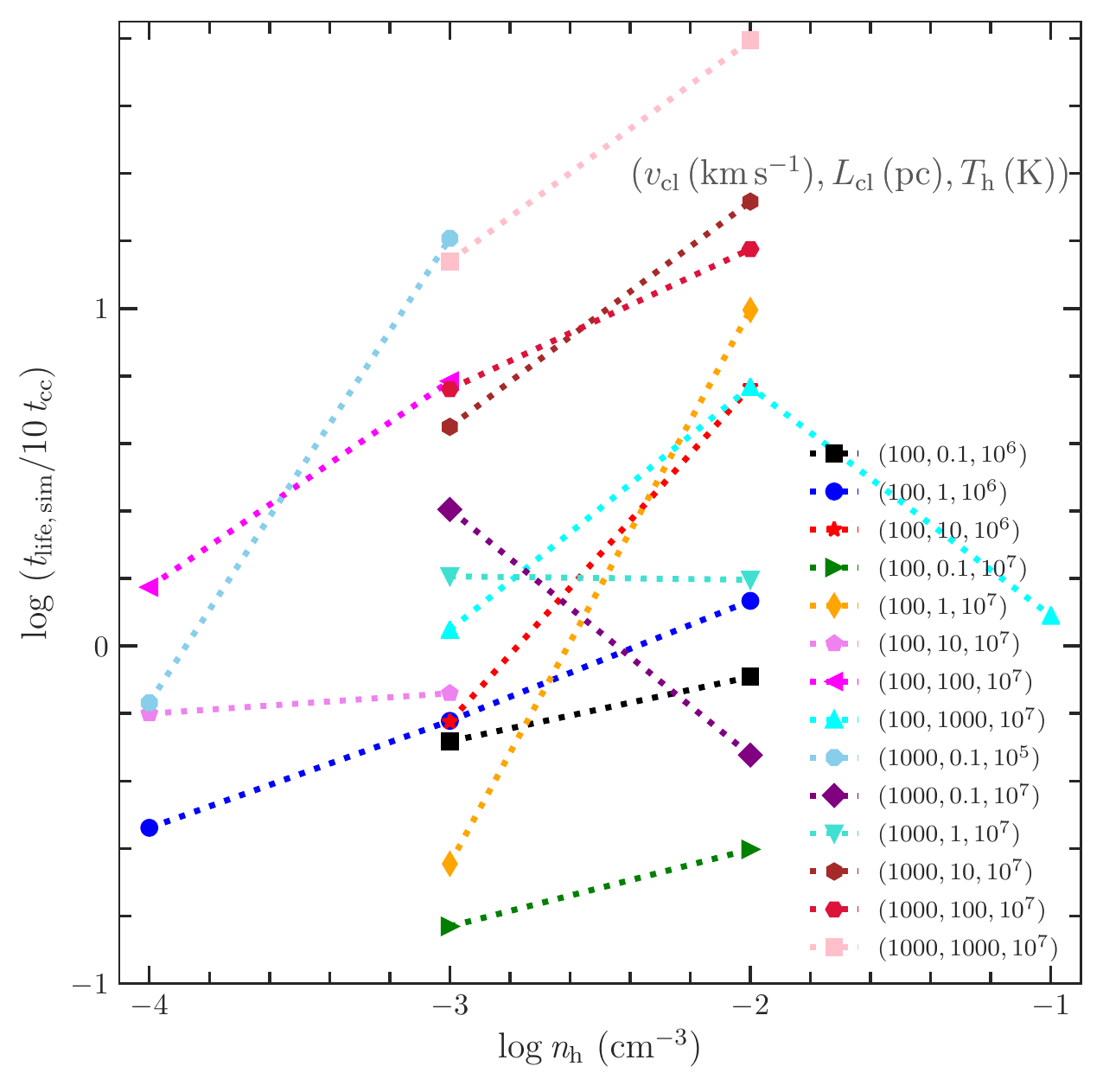}\\
\includegraphics[width=0.49\textwidth]{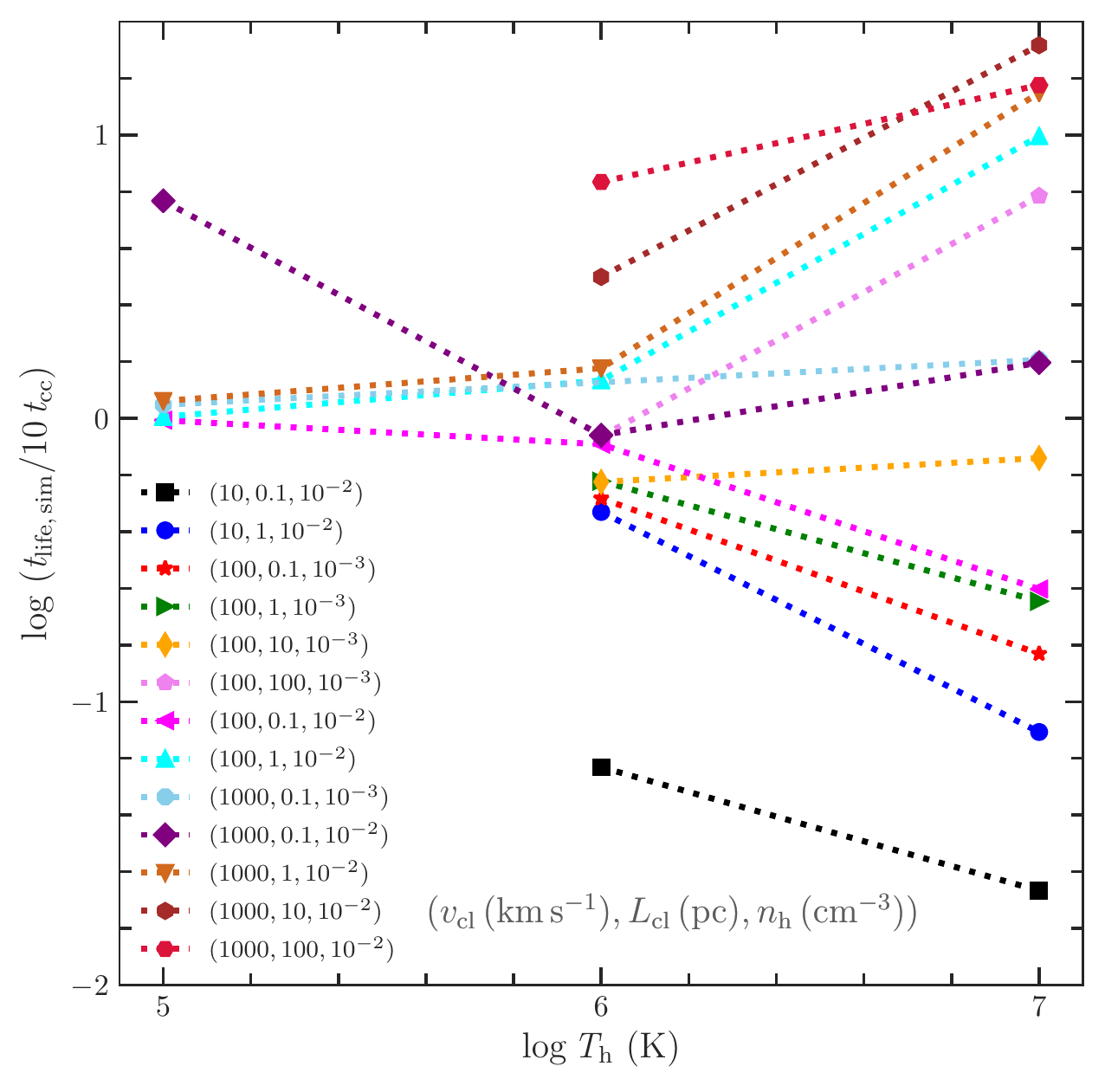}
\includegraphics[width=0.49\textwidth]{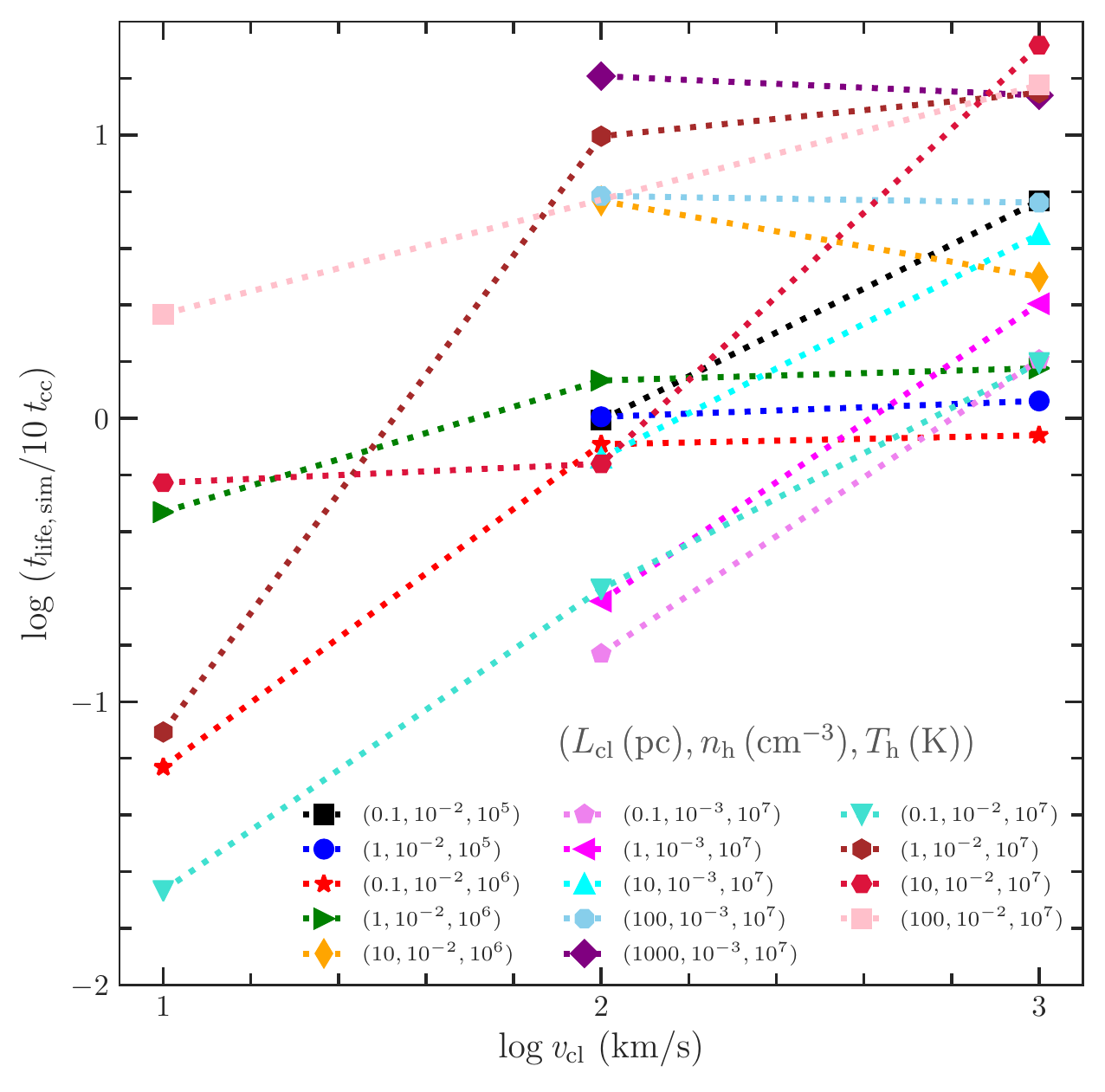}\\
    \vspace{-0.25cm}
    \caption{Simulated cloud ``lifetimes'', $t_{\rm life,\,sim}$ (in units of ten cloud-crushing time, 10\,${t_{\rm cc}}$) v.s. different initial conditions: cloud size $L_{\rm cl}$, ambient density $n_{\rm h}$, ambient temperature $T_{\rm h}$ and cloud velocity $v_{\rm cl}$. Dotted lines connect simulations that have one varying parameter but otherwise identical initial conditions. In units of ${t_{\rm cc}}$, the cloud lifetime has a weak dependence on $T_{\rm h}$, modestly increases with $L_{\rm cl}$ and $n_{\rm h}$ (i.e., cloud $N_{\rm H}$), and a slightly stronger dependence on $v_{\rm cl}$. These dependencies are captured in the scaling of $t_{\rm life,\,pred}$ with $\tilde{f}$ in Eq. \ref{eqn:scaling.relation}. Note that we factor out ${t_{\rm cc}}$ because it is the dominant effect here: our most extreme cases differ by factors of $\sim 10^8$ in their absolute lifetimes or values of ${t_{\rm cc}}$ (see e.g., Figure \ref{fig:criteria}); the ``residuals'' here, while still large ($\sim$ 1 dex), are much smaller.
    \label{fig:lifetimevsparams}}
\end{figure*}

\begin{figure}
\includegraphics[width=0.48\textwidth]{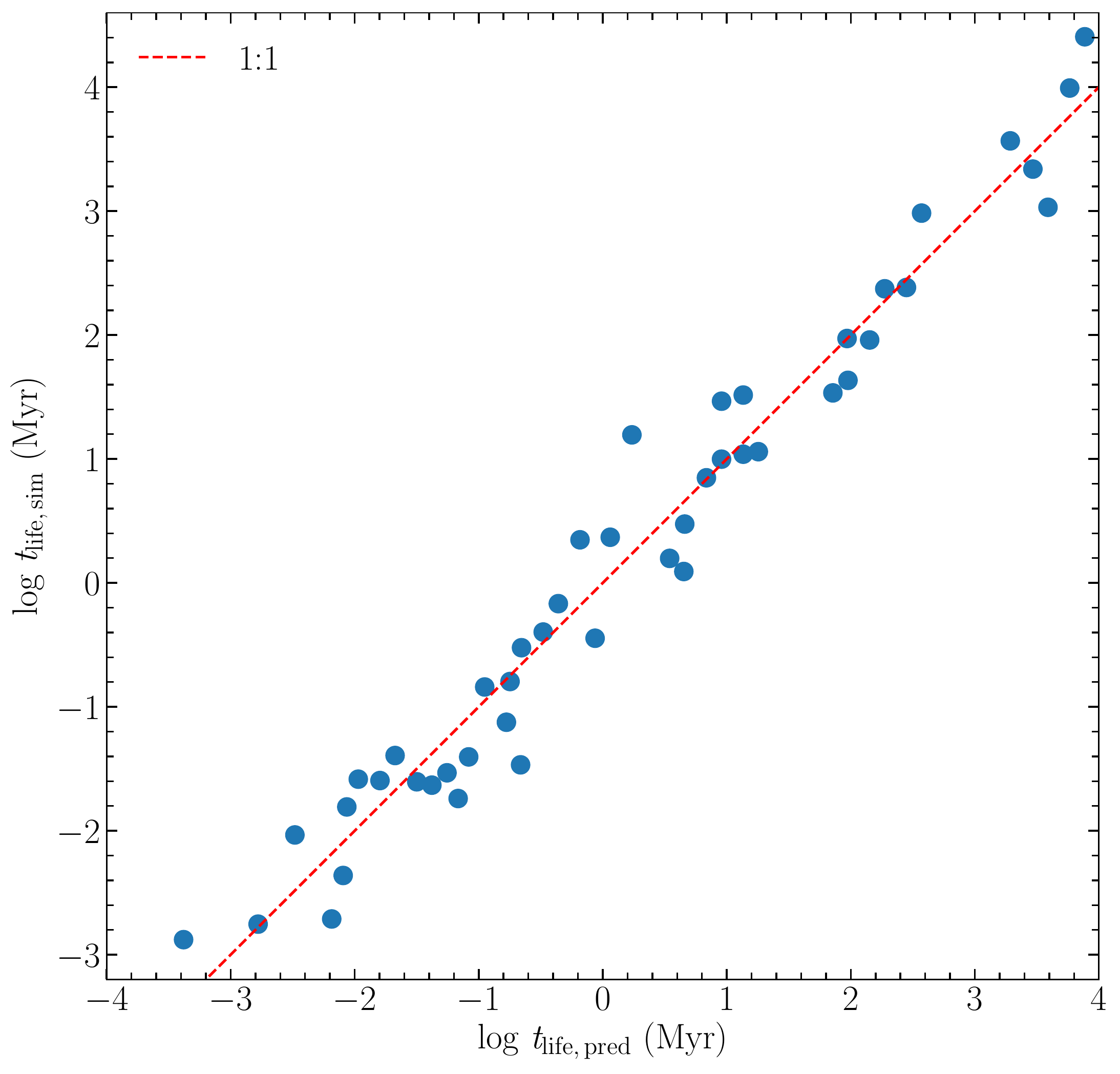} 
    \vspace{-0.25cm}
    \caption{Cloud lifetimes measured in simulations ($t_{\rm life,\,sim}$) versus the ``predicted'' lifetimes ($t_{\rm life,\,pred}$) from a simple multi-variable power-law fit to $t_{\rm life}$ versus $L_{\rm cl}$, $n_{\rm h}$, $T_{\rm h}$, and $v_{\rm cl}$, given in Eq.~(\ref{eqn:scaling.relation}). Given a dynamic range $\sim 10^{8}$ in absolute cloud lifetimes, the simulations can be remarkably well-fit by a power law of the form $t_{\rm life,\,pred} \approx 10\,t_{\rm cc}\,\tilde{f}$ with $\tilde{f}\sim L_{1}^{0.3}\,n_{0.01}^{0.3}\,v_{100}^{0.6}$ (so $\tilde{f}$ encompasses all deviations from the cloud-crushing scaling). 
    \label{fig:calibration}}
\end{figure}

Before discussing physics, it is helpful to analyze our full simulation set to understand how the cloud lifetime varies with different parameters. Given the non-scale-free nature of the physical effects we include, there is not an obvious set of dimensionless parameters with which to fit the data, so we opt to simply use the physical parameters $L_{\rm cl}$, $n_{\rm h}$, $T_{\rm h}$, and $v_{\rm cl}$. Figure \ref{fig:lifetimevsparams} shows that how the cloud lifetimes, normalized by classical cloud-destruction time $t_{\rm cc}$, scale with each of these four parameters. We perform a multi-variable log-linear fitting to these four parameters, and find that predicted lifetime scales as approximately, 
\begin{equation}\label{eqn:scaling.relation}
\begin{split}
& {t_{\rm life,\,pred}}\approx{10\,t_{\rm cc}} \tilde{f} \\
& \tilde{f} \equiv (0.9\pm0.1)\,L_{\rm 1}^{0.3}\,n_{\rm 0.01}^{0.3}\,T_{\rm 6}^{0.0}\,v_{\rm 100}^{0.6}
\end{split}
\end{equation}
where $L_{1} \equiv L_{\rm cl}/1\,$pc and $n_{0.01} \equiv n_{\rm h}/0.01\,$cm$^{-3}$. The 1-$\sigma$ values of the power-law dependences on [$L_{\rm cl}$, $n_{\rm h}$, $T_{\rm h}$, $v_{\rm cl}$] are [0.3$\pm$0.1, 0.3$\pm$0.1, 0.0$\pm$0.1, 0.6$\pm$0.1]. This fit is plotted in Figure \ref{fig:calibration}. For clouds with $v_{\rm h}$ > 10\,${\rm km\,s^{-1}}$, and for clouds in a cooler ambient medium with $T_{\rm h}$ = 10$^5$\,K, the dependence of ${t_{\rm life,\,pred}}/{t_{\rm cc}}$ on $v_{\rm h}$ is much weaker. This is discussed further in \S\ref{sec:regime:destruction:conduction} below.
 
Given the complex and non-scale-free physics involved in our default simulations, the fit (Eq. \ref{eqn:scaling.relation}) is remarkably universal. In particular, it is rather surprising that by simply assuming a separable power law in each variable, we have almost directly reproduced the classical cloud-crushing time, aside from the small correction factor $\tilde{f}$. We now discuss the reason for this universality by discussing in turn the effects that different physics have on the cloud-crushing process. These effects are shown graphically in Figure~\ref{fig:diffphys_snapshots}, showing a cloud in the process of being crushed, as we successively add physics to the pure hydrodynamical simulation (far left) in the form of (from left to right) cooling, magnetic fields, conduction, viscosity, self-shielding, and self-gravity.

\subsubsection{Effect of Radiative Cooling}
\label{sec:regime:destruction:cooling}

\begin{figure*}
\includegraphics[width=1.0\textwidth]{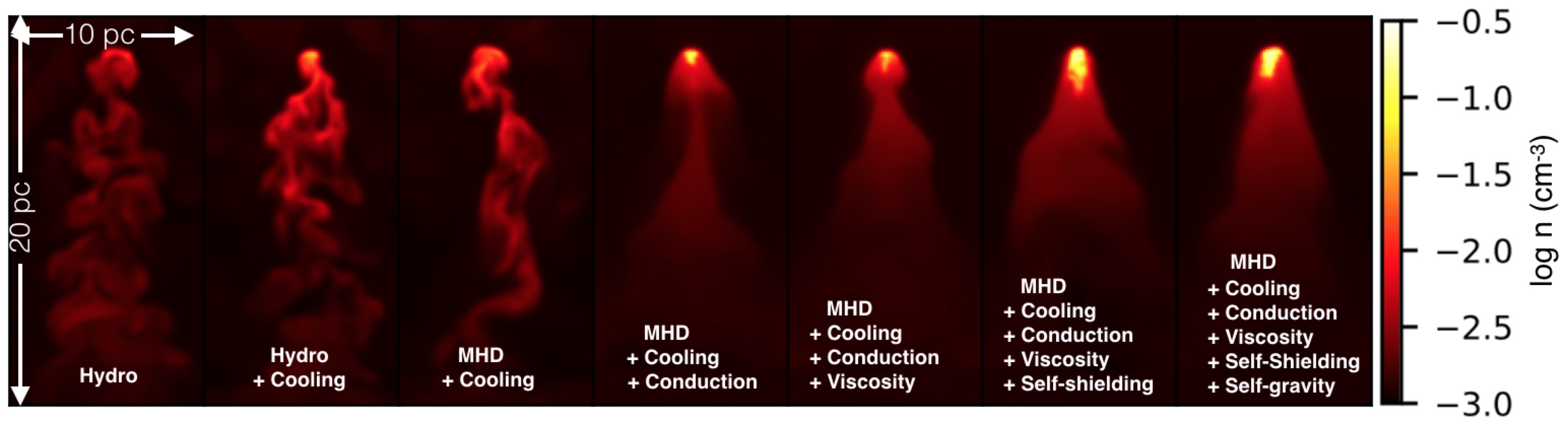}
    \vspace{-0.25cm}
    \caption{Sliced density maps for a cloud in the ``classical cloud destruction'' regime ($T_{\rm h}$ = 10$^6$\,K, $v_{\rm cl}$ = 100\,${\rm km\,s^{-1}}$, $n_{\rm h}$ = 10$^{-3}\,{\rm cm^{-3}}$, $L_{\rm cl}$ = 1\,pc), with each panel from left to right showing a simulation that includes additional physical effects (at the same physical time, 0.3 Myr). From left to right we show: Hydro = ideal hydrodynamics; Hydro + Cooling = ideal hydrodynamics + radiative cooling, etc. Our default physics set is MHD + Cooling + Conduction + Viscosity. The cloud mass evolution curves for the same set of simulations are shown in Figure \ref{fig:diffphys}.\label{fig:diffphys_snapshots}}
\end{figure*}

\begin{figure} 
\includegraphics[width=0.49\textwidth]{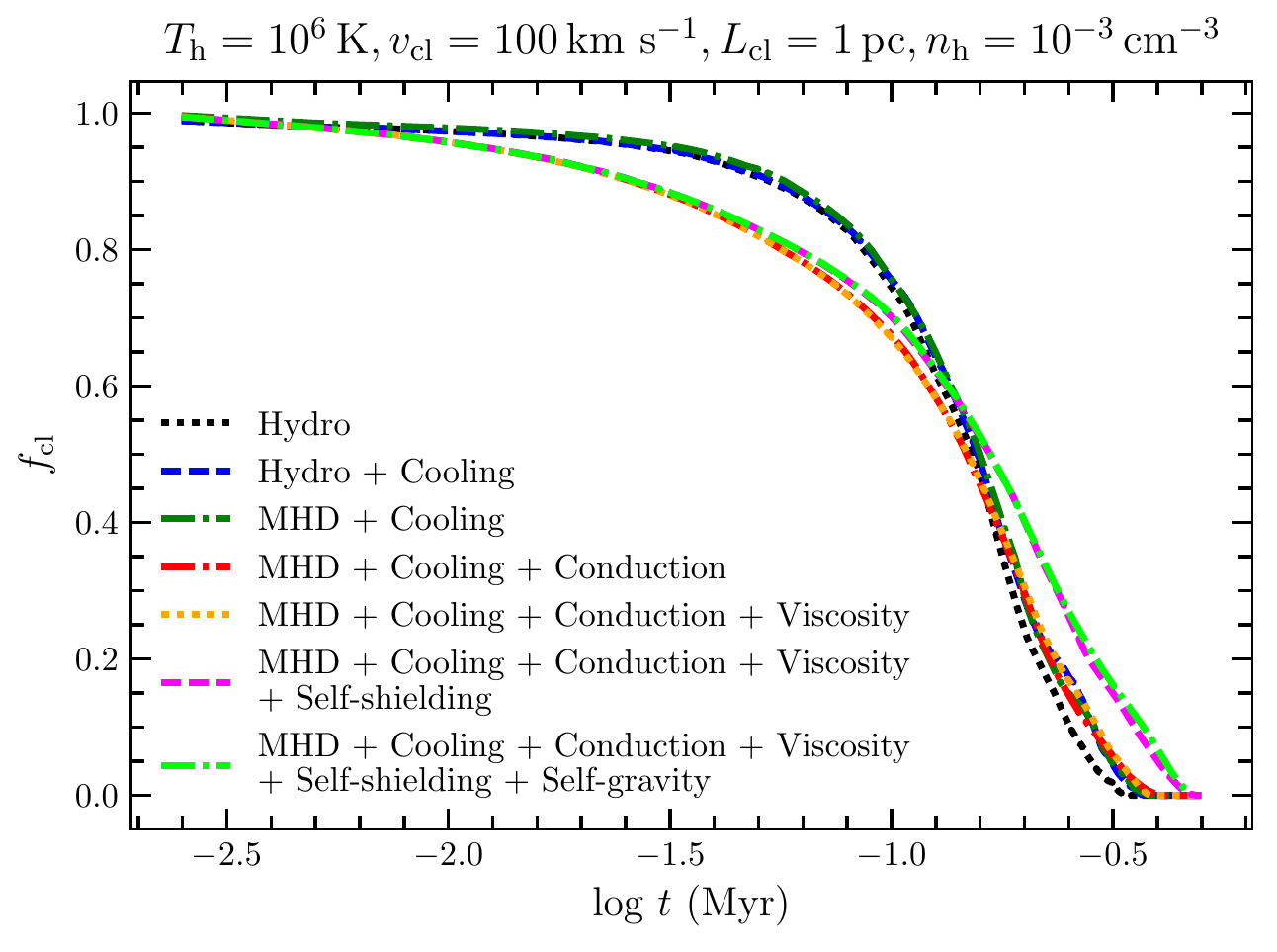}
    \vspace{-0.25cm}
    \caption{Evolution of the normalized cloud mass, $f_{\rm cl}$ (defined in Figure \ref{fig:mcloud_definition}) versus time, for the simulations shown in Figure \ref{fig:diffphys_snapshots} ($T_{\rm h}$ = 10$^6$\,K, $v_{\rm cl}$ = 100\,${\rm km\,s^{-1}}$, $n_{\rm h}$ = 10$^{-3}\,{\rm cm^{-3}}$, $L_{\rm cl}$ = 1\,pc) with different physics included (labeled as in Figure \ref{fig:diffphys_snapshots}). The cloud mass versus time is remarkably similar across these runs, given the different physics and morphologies in Figure \ref{fig:diffphys_snapshots}.
    \label{fig:diffphys}}
\end{figure}

Radiative cooling has a modestly significant effect on cloud lifetime, as discussed in previous works (see, e.g., Section 5.3 of \citealt{Klein1994}). The basic effect of cooling on gas is to soften its equation of state (lower $\gamma$), which effectively renders the cloud more compressible \citep{2015ApJ...805..158S}. This makes the cloud more strongly crushed in the direction transverse to the flow, forming a thinner, denser filament with a smaller cross section. Although KH instabilities can grow more violently on this thinner cloud than for an adiabatic cloud because it moves faster with respect to the hot medium (due to its smaller drag), the net effect is for the cloud to survive modestly longer than an equivalent cloud with no cooling due to its higher density. This behavior is nicely illustrated by the comparison of the black and blue curves in Figure \ref{fig:diffphys}. Moreover, as shown in the left two panels of Figure \ref{fig:diffphys_snapshots}, cooling can also enhance the formation of smaller, denser cloudlets in the wake \citep{2018MNRAS.473.5407M}. This effect, however, can be suppressed by magnetic fields (Figure \ref{fig:diffphys_snapshots}, see also \citealt{Gronnow2018}). Detailed analyses of the cloudlet properties have been carried out in several recent works (e.g., \citealt{Sparre2019}). 

\begin{figure}  
\includegraphics[width=0.49\textwidth]{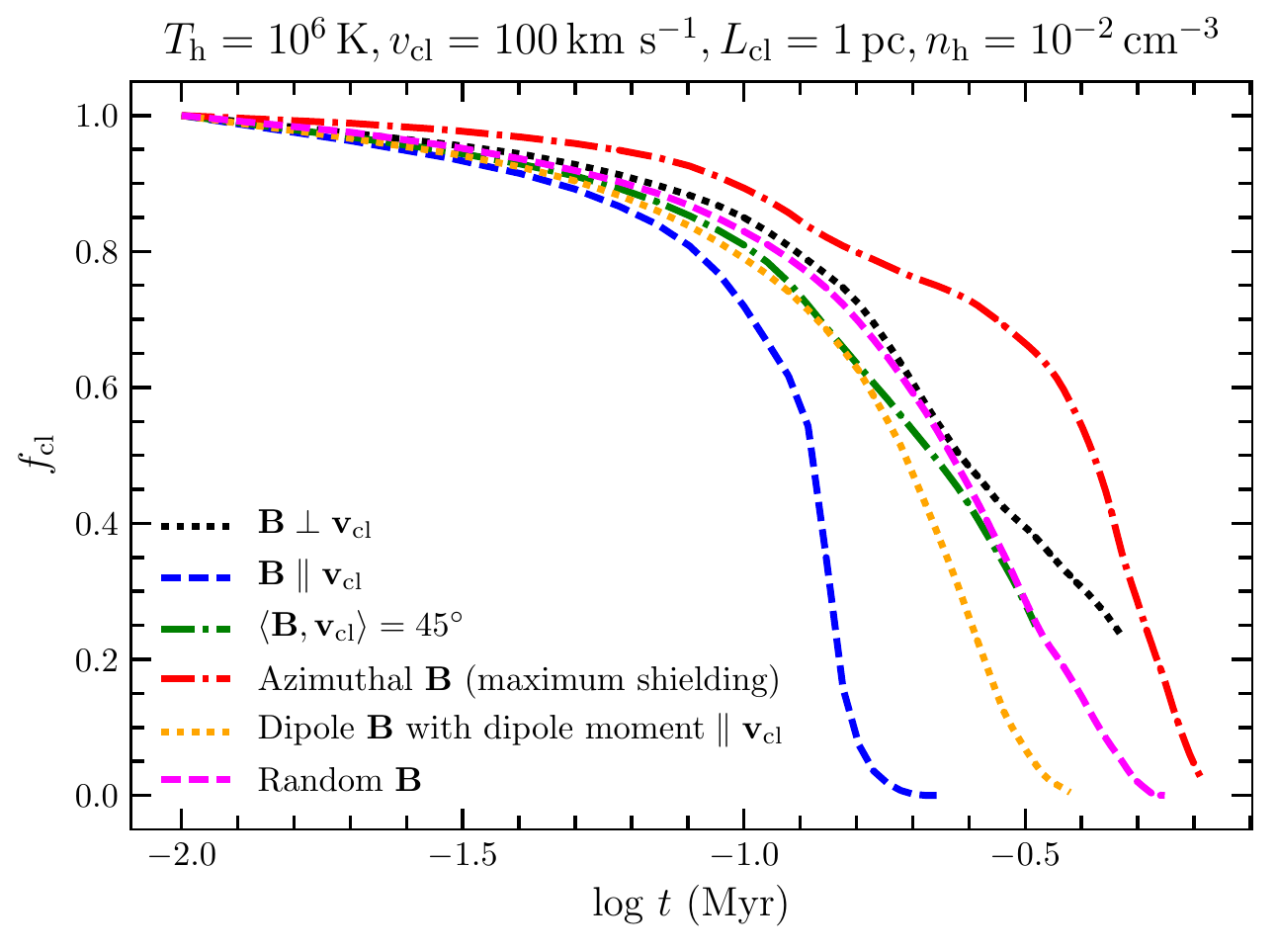}
    \vspace{-0.25cm}
    \caption{Evolution of the normalized cloud mass, $f_{\rm cl}$ (defined in Figure \ref{fig:mcloud_definition}) versus time, for otherwise identical initial conditions ($T_{\rm h}$ = 10$^6$\,K, $v_{\rm cl}$ = 100\,${\rm km\,s^{-1}}$, $n_{\rm h}$ = 10$^{-2}\,{\rm cm^{-3}}$, $L_{\rm cl}$ = 1\,pc) with different magnetic field configurations. We can see that when the magnetic field is aligned with the relative velocity ($\rm {\bf B} \parallel {\bf v}_{\rm cl}$), the cloud mass decreases most rapidly. For the azimuthal configuration (looped magnetic fields inside the cloud plus $\rm {\bf B} \perp {\bf v}_{\rm cl}$ outside the cloud, which produces maximal shielding to conduction), the cloud mass decreases most slowly. In all other cases, the magnetic field configuration does not have a large effect on the mass evolution: the lifetimes are identical to within a factor of $<2$.
    \label{fig:diffB}}
\end{figure}

\subsubsection{Effect of Magnetic Fields}
\label{sec:regime:destruction:bfield}

Magnetic fields can modify cloud destruction in two qualitatively distinct ways: (1) dynamically (via magnetic pressure or tension), or (2) by suppressing conduction/viscosity. 

Regarding (1), the magnetized ``cloud-crushing'' problem without cooling, conduction, or viscosity is well-studied \citep[see][and references therein]{1994ApJ...433..757M, 1996ApJ...473..365J, 2008ApJ...680..336S}; for very strong fields within or surrounding the cloud such that magnetic pressure is comparable to ram pressure (i.e., $P_{\rm B} \gtrsim P_{\rm ram} \sim \rho\,v_{\rm cl}^{2}$, or $\beta \lesssim \mathcal{M}_{\rm h}^{-2}$), cloud destruction is strongly suppressed. While $\beta \lesssim 1$ is common in very cold (e.g., molecular) gas in the ISM, in the warm and hot CGM realistic estimates of $\beta$ range from $\sim 10^{2}-10^{9}$ \citep[see][]{2017MNRAS.471..144S, Martin2018, Hopkins19}, viz., the direct dynamical effects of the fields are negligible. Alternatively, it has been proposed that a strong field could build up via ``magnetic draping'' \citep{2007PhR...443....1M}, wherein the cloud ``sweeps up'' field lines oriented perpendicular to ${\bf v}_{\rm cl}$, compressing the field leading the cloud and increasing $|{\bf B}|$. \citet{1999ApJ...517..242M} define the ``draping time,\footnote{We emphasize that the context in which draping was originally proposed referred to much larger structures, namely ``bubbles'' and jets emanating from AGN in the CGM of massive halos/clusters, which have physical size scales $\sim 10-100\,$kpc and travel $\gtrsim 100\,$kpc, vastly different from what we model here.}'' which we can turn into the equivalent length: 
\begin{equation}
L_{\rm drape} \sim \frac{\pi\,R_{\rm cl}\chi^{2/3}}{50}\,\left(\frac{P_{\rm ram}+P_{\rm therm}}{P_{\rm B}}\right)^{2/3}\approx 3\,{\rm  kpc}\,R_{\rm pc}\,(\beta_{1000}\,T_{6}\,v_{100}^{2})^{2/3}
\end{equation}
$L_{\rm drape}$ is the path length that a cloud must travel for the accumulated field to appreciably alter its destruction (assuming $P_{\rm therm} \ll P_{\rm ram}$ for supersonic clouds). However, $L_{\rm drape} $ is much longer than the length scale over which clouds are destroyed, $L_{\rm cc}\approx t_{\rm cc}\,v_{\rm cl} \approx 9 \,{\rm pc}\, R_{\rm pc} v_{100} \mathcal{M}_{\rm cl}^{-1}$. In other words, CGM magnetic fields are nowhere near sufficiently strong to dynamically suppress cloud destruction. This can be seen visually by comparing the second and third panels of Figure \ref{fig:diffphys_snapshots} (or the relevant lines in Figure \ref{fig:diffphys}), which shows how MHD and hydrodynamic simulations remain very similar without the effects of conduction. We have also confirmed this conclusion by  re-running a subset of our simulations with plasma $\beta$ multiplied or divided by a factor of $\sim 1000$, which makes no difference to the measured lifetimes (as expected, since they remain in the weak-field limit).

However, regarding (2), even a very weak field is sufficient to suppress perpendicular conduction, viscosity (typically the perpendicular transport coefficients are suppressed by $\sim$ $\lambda_{e, \rm gyro}/\lambda_{e,\,\rm h}$ $\sim 10^{-8}$) and hydrodynamic instabilities \citep{Dursi2008, Banda-Barragan2016, Banda-Barragan2018}.
In this case the field {\em geometry} is what matters, while the field strength is irrelevant. In Figure \ref{fig:diffB}, we therefore explore a series of simulations of one of our typical cloud-destruction cases, varying the initial field geometry. In general, the magnetic field configuration does not have a strong effect on the evolution of cloud mass. This is not surprising, as draping can rearrange the geometry of the magnetic field around the cloud to similar configurations and yield similar amount of suppression of conduction, viscosity and instabilities, regardless of the initial field geometry (note that the arguments of \S\ref{sec:regime:destruction:conduction} below suggest that conduction plays only a secondary role anyway). However, in several extreme cases, such as when the magnetic field is aligned with the relative velocity ($\rm {\bf B} \parallel {\bf v}_{\rm cl}$),  we do see a more rapid decrease in the cloud mass as there is essentially no draping. In contrast, with an azimuthal field configuration (looped magnetic fields inside the cloud plus $\rm {\bf B} \perp {\bf v}_{\rm cl}$ outside the cloud), the cloud mass decreases most slowly, indicating that the field can shield the cloud particularly efficiently in this case\footnote{Note that in the ``cloud growing'' regime, transverse magnetic fields can shield the cloud via draping, reduce both mixing and warm gas mass loading and prevent condensation (see \citealt{Gronnow2018}). Also note that self-contained magnetic fields can enhance clumping and reduce cloud destruction \citep{Li2013, McCourt15, Banda-Barragan2018}. We defer a detailed study of these effects to future work.}(see also \citealt{Li2013, Banda-Barragan2016, Gronnow2017}).

\subsubsection{Effect of Conduction}
\label{sec:regime:destruction:conduction}

The influence of conduction on isolated, undisturbed clouds (i.e., those without an impinging wind) has been studied by \citet{Cowie1977,1977ApJ...215..213M,Balbus1982}. For the range of temperatures relevant to our study ($10^{5}\,{\rm K}\lesssim T_{\rm h}\lesssim 10^{7}\,{\rm K}$) the conclusion of these papers is that cloud evaporation/condensation is controlled by the saturation parameter\footnote{We define $\sigma_0$ to match the numerical value given of $\sigma_0$ in   \citet{1977ApJ...215..213M}, which leads to a slightly different definition in terms of $\lambda_{e,\,\rm h}/R_{\rm cl}$ compared to \citet{Cowie1977} because of a different definition of $\lambda_{e,\,\rm h}$.}
\begin{equation}
\sigma_{0}\approx 3.2 \frac{\lambda_{e,\,\rm h}}{R_{\rm cl}} \approx 0.4\frac{T_{6}^{3}}{\langle n_{\rm cl} \rangle R_{\rm pc}} \approx T_{6}^{3} \left(\frac{N_{\rm H}}{1.2\times 10^{18}\, {\rm cm}^{-2}} \right)^{-1}\label{eq:cowie.sigma.parameter}
\end{equation}
For small values of $\sigma_{0} \lesssim 0.01$ (large clouds), the cooling of the hot material onto the cloud is sufficiently rapid that the cloud condenses. The necessary size of such clouds ($N_{\rm H}\gtrsim 1.2\times 10^{20}\, T_{6}^{3}\,{\rm cm}^{-2}$) corresponds, within an order of magnitude, to the ``growing-cloud'' regimes discussed in \S\ref{sec:results:regimes:grav.shield}--\S\ref{sec:results:regimes:growth} (the cloud sizes required for growth in the crushed problem are slightly larger, which intuitively makes sense given they are being actively ripped apart by the wind). On the other side, large values of $\sigma_{0}\gtrsim \chi$ correspond to the smallest clouds discussed in \S\ref{sec:results:regimes:tiny}, which are immediately evaporated by hot electrons penetrating throughout the entire cloud \citep{Balbus1982}. Thus, effectively all of our clouds in the ``classical cloud destruction'' regime lie in the range $0.01 \lesssim \sigma_{0}\lesssim \chi$, which, in the absence of the hot wind would slowly evaporate into the ambient medium. As shown by \citet{1977ApJ...215..213M}, the conductive heat flux that evaporates the cloud is in the unsaturated regime for clouds with $\sigma_{0}\lesssim 1$, while the heat flux is saturated for $\sigma_{0}\gtrsim 1$.

To make further progress, let us compare the cloud evaporation timescale to the cloud-crushing time. In the $\sigma_{0}\lesssim 1$ regime, \citet{Cowie1977} compute the mass-loss rate by solving the hydrodynamic equations in spherical geometry, deriving the evaporation time of the cloud as (setting ln $\Lambda_{D}$ = 30) \begin{equation}
t_{\rm evap}\approx 30\,{\rm Myr}\,n_{\rm 0.01} R_{\rm pc}^{2}T_{6}^{-5/2}
\end{equation}
In the $\sigma_{0}\gtrsim 1$ regime, where the heat flux is saturated,  one can derive the evaporation
time by comparing the rate at which energy is transferred to the cold cloud due to the saturated heat flux, \begin{equation}
\dot E = 4\pi R_{\rm cl}^2\,q_{\rm sat}\approx 4\pi \alpha R_{\rm cl}^2\,n_{\rm h}\,c_{s,\,e,\,\rm h}k_{B}\,T_{\rm h}
\end{equation}
(here $\alpha\approx0.3$ is chosen to match Eq. \ref{eqn:kappa.gizmo}), to the total energy required to evaporate the cloud by heating it up to the hot-medium temperature, \begin{equation}
E\approx \frac{4}{3}\pi R_{\rm cl}^3\,n_{\rm cl}\,k_{B}\,T_{\rm h}
\end{equation}
(A more complicated approach in \citet{Cowie1977} gives a similar estimate; see their Eq. 64).
Because the heat flux is effectively given by the minimum of the unsaturated and saturated values (see Eq. \ref{eqn:kappa.gizmo}), the time for the cloud to evaporate is the maximum of the unsaturated and saturated estimates, or 
\begin{equation}
\frac{t_{\rm evap}}{t_{\rm cc}} \approx \max\left\{ 2 \mathcal{M}_{\rm h}n_{\rm cl}L_{\rm pc} T_{6}^{-5/2},\,0.3 \mathcal{M}_{\rm h}T_{6}^{1/2} \right\}\label{eq:tevap.for.conduction}
\end{equation}
Note that the saturated (right-hand) expression is simply $\approx v_{\rm cl}/(300\,{\rm km\,s}^{-1})$. 

We see that across the range of parameters surveyed, $t_{\rm evap}/t_{\rm cc}$ ranges from much larger than 1 for large clouds in fast winds, to somewhat less than 1 for  smaller clouds. What will be the effect of this evaporation on the cloud-crushing  process? For $t_{\rm evap}/t_{\rm cc}\ll1$ we expect the cloud to behave effectively as it would in the absence of a wind, evaporating rapidly into the hot medium. On the other hand, when $t_{\rm evap}/t_{\rm cc}\gtrsim1$ the evaporation has only a minor effect on the cloud lifetime, because it is crushed by the wind before the heat flux has much of an effect (the static approach of \citealt{Cowie1977} also becomes highly questionable in such a strongly perturbed cloud). There does, however, seem to be a reasonably significant effect on the cloud morphology, which is evident in the change between the third and fourth panels of Figure \ref{fig:diffphys_snapshots} (see also \citealt{2016ApJ...822...31B}). This type of behavior, which occurs at $t_{\rm evap}/t_{\rm cc}\sim 1$, seems to be related to the fast creation of a conductive boundary layer, which causes an inwards pressure on the cloud due to the outflow of hot material from its outer edges. This compresses the cloud and increases its density, which sometimes has the effect of modestly increasing the cloud lifetime. Indeed, if we make the gross approximation that the mass is lost from the cloud with an outflow velocity that is approximately the ion sound speed (since the ions will be heated by the impinging hot electrons to approximately $T_{\rm h}$), one finds that the ratio of the inwards pressure due to the outflow ($P_{\rm evap} \approx \dot{m}\,v_{\rm out}/(4\pi R^2) \approx m\,v_{\rm out}/(4\pi R^2 t_{\rm evap})$) to the thermal pressure of the cloud ($P_{\rm cl}$) is approximately 
\begin{equation}
\frac{P_{\rm evap}}{P_{\rm cl}}\approx \min \left\{2 \frac{T_{6}^{3}}{n_{\rm cl} R_{\rm pc}}, 10 \right\}\label{eq:evaporative.pressure}
\end{equation}
where the left-hand expression is that of the unsaturated ($\sigma_{0}\lesssim 1$) regime, and the right-hand expression is that of the saturated ($\sigma_{0}\gtrsim 1$) regime. We thus see that for smaller clouds, the pressure from evaporative outflow is modestly large compared to that of the cloud, and should thus be able to cause some compression, as seen in Figure \ref{fig:diffphys_snapshots}.
 
\begin{figure}
\includegraphics[width=1.0\columnwidth]{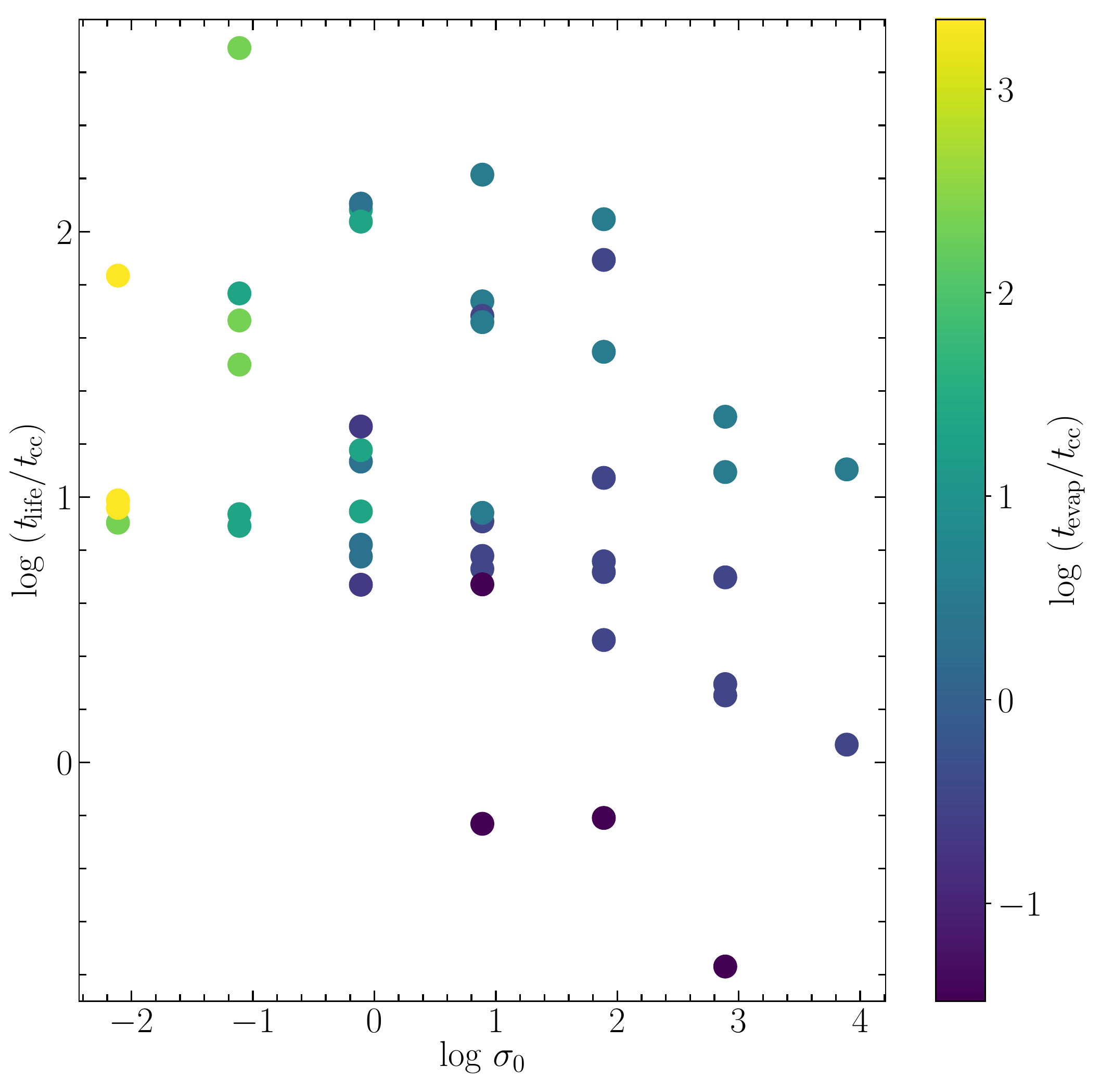}
\caption{The simulated cloud lifetimes in units of the cloud-crushing time, $t_{\rm life}/t_{\rm cc}$, v.s. the saturation parameter, $\sigma_{0}$ (Eq. \ref{eq:cowie.sigma.parameter} in \S~\ref{sec:regime:destruction:conduction}, which quantifies the strength of conduction) for clouds in the ``classical destruction'' regime. The simulations are color-coded from light yellow to dark blue with decreasing $t_{\rm evap}/t_{\rm cc}$, where $t_{\rm evap}$ is the cloud evaporation time for a non-moving cloud in a conducting medium (Eq. \ref{eq:tevap.for.conduction}). Simulations with $t_{\rm evap} \ll t_{\rm cc}$ are evaporated before cloud-crushing, explaining why $t_{\rm life} \ll t_{\rm cc}$. These clouds almost exclusively have $\sigma_{0} \gg 1$, i.e., are in the regime of saturated conduction, where $t_{\rm evap}$ $\propto$ $t_{\rm cc} v_{\rm cl}$, explaining the strong dependence of $\tilde{f}$ on $v_{\rm cl}$. While for simulations with $t_{\rm evap} \gg t_{\rm cc}$, clouds are only weakly influenced by conduction, and therefore $t_{\rm life}$ $\propto$ $t_{\rm cc}$.}
\label{fig:effect.of.conduction.all.sims}
\end{figure}

The broad ideas of the previous paragraphs are confirmed in Figure \ref{fig:effect.of.conduction.all.sims}, which plots $t_{\rm life }/t_{\rm cc}$ v.s. $\sigma_{0}$ for our full suite of simulations, with each point colored by $t_{\rm evap}/t_{\rm cc}$ from Eq. \ref{eq:tevap.for.conduction}. We see that, as expected, only those simulations with $t_{\rm evap}/t_{\rm cc}\ll1$ are destroyed significantly faster than $t_{\rm cc}$ (these are all low-velocity clouds). The lifetime of simulations with $t_{\rm evap}/t_{\rm cc}\gtrsim 1$ is mostly independent of $\sigma_{0}$, aside from a possible slight increase in lifetime for $\sigma_{0}\gtrsim 1$, which may be indicative of cloud compression due to the evaporative outflow. Finally, we note that this general framework explains our measured empirical scaling of $t_{\rm life}/t_{\rm cc}$ with a positive power of $v_{\rm cl}$ (see Eq.~\ref{eqn:scaling.relation}), because the lowest velocity clouds are quickly destroyed by saturated conduction, i.e., their $t_{\rm life}$ $\sim$ $t_{\rm evap}$ $\propto$ $t_{\rm cc} v_{\rm cl}$ (Eq. \ref{eq:tevap.for.conduction}), while those with higher velocities can live somewhat longer than $t_{\rm cc}$ due to the evaporative compression to higher densities. Meanwhile, for e.g., $T_{\rm h}=10^{5}$K, all clouds fall into the $\sigma_{0}\lesssim 1$ regime (see Eq.~\ref{eq:cowie.sigma.parameter}), where $t_{\rm evap}/t_{\rm cc}>1$ and the evaporative pressure (Eq.~\ref{eq:evaporative.pressure}) is unimportant, so we simply obtain $t_{\rm life}\propto t_{\rm cc}$.

\subsubsection{Effect of Viscosity}
\label{sec:regime:destruction:viscosity}

The effect of viscosity is in general sub-dominant to conduction. This is not surprising because conduction is controlled by the thermal velocity of hot electrons, while viscosity is controlled by the thermal velocity of ions, and the ratio of these thermal velocities (and thus the strength of conductivity and viscosity) is $(m_{i}/m_{e})^{1/2}$ $\sim$\,40, assuming each has the same temperature. Nonetheless, viscosity does provide some non-zero insulating effects as a viscous ``boundary layer'' that forms around the cloud, which drags the co-moving boundary layer and can slightly increase the cloud lifetime for some clouds. This minor effect can be seen through the comparison of the fourth and fifth panels in Figure~\ref{fig:diffphys_snapshots}.

\subsubsection{Effect of Turbulence in the Cloud or Ambient Medium}
\label{sec:regime:destruction:turbulence}

Some historical studies have argued that clouds which have initial ``turbulence'' (large density and velocity fluctuations) like GMCs in the ISM \citep[e.g.,][and references therein]{2015ApJS..217...24S} might be much more rapidly-disrupted. However, most of these studies have considered clouds with large internal turbulent Mach numbers $\mathcal{M}_{\rm cl}^{\rm turb} \equiv |\delta {\bf v}_{\rm turb}|/c_{s,\,\rm cl} \sim 10-100$, akin to GMCs (see \S~\ref{sec:intro}), e.g.,\ \citet{2015ApJS..217...24S} consider an internal 3D Mach number $\mathcal{M}_{\rm cl}^{\rm turb} \sim 9$ (or equivalently, 1D Mach number $\mathcal{M}_{\rm cl}^{\rm turb} \sim 5$), which produces nearly $\sim 1\,$dex initial rms density fluctuations. 

However, for realistic turbulent Mach numbers in the CGM, turbulence should produce much weaker effects. This is because the initial cloud temperature is $10^{4}$\,K ($c_{s,\,{\rm cl}} = 10\ {\rm km\,s^{-1}}$), as compared to $\sim 10\,$K in GMCs, and the density and temperature fluctuations only become very large for large turbulent Mach numbers ($\mathcal{M}_{\rm cl}^{\rm turb} \gg 1$), which are highly unrealistic in the CGM (e.g.,\ clouds do not have internal velocity dispersions of $\sim 100\,{\rm km\,s^{-1}}$). The turbulent Mach numbers should be even lower in the hot medium. Moreover, $\mathcal{M}_{\rm cl}^{\rm turb} \gg 1$ is not a self-consistent ``cloud'' under the conditions we consider, because it necessarily implies a turbulent ram pressure much larger than the confining gas pressure (the ``cloud'' would simply fly apart as soon as the simulation begins): in GMCs this is resolved by confinement via self-gravity, but we have already excluded this regime.

We therefore have considered a subset of simulations using initial conditions drawn from driven periodic box simulations of turbulence \citep[taken from][]{colbrook:passive.scalar.scalings}, for the cloud itself, the ambient hot medium, or both, with Mach numbers in each medium of $\sim 0.1,\,0.5,\,1$. Not surprisingly, these have little effect on the supersonic cloud-crushing process (consistent with \citealt{Banda-Barragan2018, Banda2019}). For example, for $\mathcal{M}_{\rm cl}\sim 0.1$, the initial density and pressure fluctuations are only of the order of $\sim\! 1\%$, much smaller than those introduced almost immediately by the cloud-wind interaction. We therefore do not discuss these cases in more detail.

\section{Conclusions}
\label{sec:conclusions}

In this paper, we have systematically explored the survival of cool clouds traveling through hot gas -- the so-called ``cloud crushing'' problem -- for parameters relevant to the CGM. We present a  comprehensive parameter survey, with cloud diameters from $\sim 0.01-1000\,$pc, relative velocities $\sim 10-1000\, {\rm km\,s^{-1}}$, ambient temperatures $\sim 10^{5}-10^{7}\,$K and ambient densities $\sim 10^{-4} - 10^{-1}\, {\rm cm^{-3}}$. We study the effects of a range of physics, including radiative cooling, anisotropic conduction and viscosity, magnetic fields, self-shielding and self-gravity. We identify several unique regimes, which give rise to qualitatively different behaviors, including collapse, growth, expansion, shredding, and evaporation. For mid-sized clouds, those in the ``classical cloud destruction'' regime, we also quantify the cloud lifetime as a function of parameters across the broad range of initial conditions. We reach a number of important conclusions, including: 
\begin{enumerate} 

\item{Clouds which are {\em initially} self-gravitating/Jeans-unstable, or self-shielding to molecular/low-temperature metal-line fine-structure cooling and thus able to cool to temperatures $T \ll 1000\,$K, will fragment and form stars before they are disrupted. For  clouds that are initially Jeans-stable and non-shielding, these effects can be neglected. This transition occurs when the cloud exceeds large, DLA-like column densities (Eq.~\ref{eqn:NH.jeans}, \ref{eqn:NH.shield}).}

\item{In an ambient medium  where the ``diffuse'' gas cooling time is shorter than  the time for diffuse gas to cross  the cloud ($\sim  R_{\rm cl}/v_{\rm cl}$), pressure-confinement of the cloud cannot effectively operate and the cloud-crushing problem is ill-posed. In hotter medium ($T_{\rm h} \gtrsim 10^{6}\,$K) this only occurs at high enough column densities such that the cloud would already be self-gravitating; while in cooler ambient halos ($T_{\rm h} < 10^{6}$\,K), which are generally not able to sustain a ``hot halo'' in  quasi-hydrostatic equilibrium, even clouds with more modest column densities $N_{\rm H} \gtrsim  10^{18}\,{\rm cm^{-2}}$ can reach this regime  (see Eq.~\ref{eqn:NH.cool}).}

\item{If the expected destruction time of a cloud through shocks and fluid mixing (cloud crushing) is longer than the cooling time of the swept-up material in the shock front leading the cloud, the cloud can grow in time, rather than disrupt \citep{Gronke18}. The cooling of the shock front material adds to the cloud mass (with the growth time simply being the timescale to ``sweep up'' new mass), faster than instabilities can disrupt the cloud, and the cloud acts more like a seed for the thermal instability. This can occur at column densities well below the self-gravity/shielding/ambient medium rapid cooling thresholds above (see Eq.~\ref{eqn:NH.grow}).}

\item{If we restrict to clouds {\em below} the sizes/column densities of the above thresholds, and {\em above} the size/column density  where they become smaller than the penetration length of hot electrons into the cloud ($N_{\rm H} \gtrsim 10^{16}\,{\rm cm^{-2}}\,T_{6}^{2}$; Eq.~\ref{eqn:NH.small}), then we find that the clouds are indeed disrupted and mixed by a combination of instabilities, shocks, and conduction. Remarkably, the cloud lifetimes can be  well fit by a single power law similar to the classical ``cloud-crushing'' scaling for the pure hydrodynamic problem, albeit with a larger normalization and a secondary dependence on the ambient temperature and velocity, which is introduced by the combination of cooling and conduction. We develop simple analytic scalings to understand how this modification to the scaling arises.}

\item{Braginskii viscosity, turbulent density/velocity fluctuations in the cloud, and magnetic field geometry and strength have relatively weak effects on cloud lifetimes and do not qualitatively alter our conclusions. Viscous effects tend to  be sub-dominant to conduction because of the relative scaling of ion and  electron mean-free-paths in the CGM (although we caution that our model assumes equal ion and electron temperatures). Turbulent effects are weak for {realistic} initial cloud turbulence, because  CGM clouds, {unlike} GMCs in the ISM, cannot be highly supersonic (this would require {\em internal} turbulent Mach numbers in the cloud $\gg\,$1). This implies that the initial density fluctuations in the cloud are quite small. Magnetic field strength has little effect because the CGM plasma has $\beta \gg 1$ (i.e.,\ magnetic pressure is much weaker than thermal pressure, which is yet smaller than the ram pressure) and the distance clouds would have to travel to acquire dynamically important fields via ``draping'' is much longer than the length over which they are destroyed. Field geometry has some effect, by suppressing thermal conduction in the directions perpendicular to the field. However, we show the net effect of the field geometry is minor  for most plausible geometries ($\sim 10\%$ in $t_{\rm life}$) and even the most extreme favorable/unfavorable field geometries produce only a factor of $\sim \!2$ systematic change in cloud lifetimes.}

\end{enumerate}

We caution that there are still a number of caveats to this study. There remain a number of simplifications in the physics included in our model  (Eqs.~\ref{eqn:momentum}--\ref{eqn:e.defn}), which may be important for some regimes. The most important of these is likely the assumption of equal electron and ion temperatures, even in the presence of strong conduction and cooling on scales approaching the electron mean free path. Indeed, because the timescale for ions to collisionally equilibrate with electrons is $\sim\!m_i/m_e$ times the electron-electron  collision timescale, regions with large (saturated) electron heat fluxes may also have $T_e\gg T_i$ or $T_i\gg T_e$. Unfortunately, tackling this issue in detail is difficult and computationally demanding even in simplified setups \citep[see, e.g.,][]{Kawazura2019}, and is well beyond current computational capabilities for a highly inhomogeneous problem such as cloud crushing. On fluid scales, there are also significant uncertainties that arise from our basic numerical setup, which we have intentionally restricted to be rather idealized. Potential complications that might be relevant and interesting to study in future work include lack of pressure equilibrium in the cool gas (as could arise from, e.g.,\ supersonic turbulence, \citealt{Banda-Barragan2018}), the effect of stratification of the ambient medium, and the  interaction with scales that are not resolved in our simulations here \citep[see, e.g., ][]{2018MNRAS.473.5407M}. However, in view of the simple physical arguments that have supplemented most of the main conclusions of this paper (see above), it seems unlikely that these effects would cause significant qualitative changes to our main results. 

\acknowledgments{Support for ZL and co-authors was provided by an Alfred P. Sloan Research Fellowship, NSF Collaborative Research Grant \#1715847 and CAREER grant \#1455342, and NASA grants NNX15AT06G, JPL 1589742, 17-ATP17-0214. JS acknowledges the support of the Royal Society Te Ap\=arangi  through a Rutherford Discovery Fellowship RDF-U001804 and Marsden Fund grant UOO1727. Numerical calculations were run on the Caltech compute cluster ``Wheeler,'' allocations from XSEDE TG-AST130039 and PRAC NSF.1713353 supported by the NSF, and NASA HEC SMD-16-7592.}

\bibliography{cloud_crushing_cgm}

\begin{thebibliography}{}
\makeatletter
\relax
\def\mn@urlcharsother{\let\do\@makeother \do\$\do\&\do\#\do\^\do\_\do\%\do\~}
\def\mn@doi{\begingroup\mn@urlcharsother \@ifnextchar [ {\mn@doi@}
  {\mn@doi@[]}}
\def\mn@doi@[#1]#2{\def\@tempa{#1}\ifx\@tempa\@empty \href
  {http://dx.doi.org/#2} {doi:#2}\else \href {http://dx.doi.org/#2} {#1}\fi
  \endgroup}
\def\mn@eprint#1#2{\mn@eprint@#1:#2::\@nil}
\def\mn@eprint@arXiv#1{\href {http://arxiv.org/abs/#1} {{\tt arXiv:#1}}}
\def\mn@eprint@dblp#1{\href {http://dblp.uni-trier.de/rec/bibtex/#1.xml}
  {dblp:#1}}
\def\mn@eprint@#1:#2:#3:#4\@nil{\def\@tempa {#1}\def\@tempb {#2}\def\@tempc
  {#3}\ifx \@tempc \@empty \let \@tempc \@tempb \let \@tempb \@tempa \fi \ifx
  \@tempb \@empty \def\@tempb {arXiv}\fi \@ifundefined
  {mn@eprint@\@tempb}{\@tempb:\@tempc}{\expandafter \expandafter \csname
  mn@eprint@\@tempb\endcsname \expandafter{\@tempc}}}

\bibitem[\protect\citeauthoryear{{Armillotta}, {Fraternali}, {Werk},
  {Prochaska}  \& {Marinacci}}{{Armillotta} et~al.}{2017}]{2017MNRAS.470..114A}
{Armillotta} L.,  {Fraternali} F.,  {Werk} J.~K.,  {Prochaska} J.~X.,
  {Marinacci} F.,  2017, \mn@doi [\mnras] {10.1093/mnras/stx1239}, \href
  {https://ui.adsabs.harvard.edu/\#abs/2017MNRAS.470..114A} {470, 114}

\bibitem[\protect\citeauthoryear{{Balbus} \& {McKee}}{{Balbus} \&
  {McKee}}{1982}]{Balbus1982}
{Balbus} S.~A.,  {McKee} C.~F.,  1982, \mn@doi [\apj] {10.1086/159581}, \href
  {https://ui.adsabs.harvard.edu/abs/1982ApJ...252..529B} {252, 529}

\bibitem[\protect\citeauthoryear{{Banda-Barrag{\'a}n}, {Parkin}, {Federrath},
  {Crocker}  \& {Bicknell}}{{Banda-Barrag{\'a}n}
  et~al.}{2016}]{Banda-Barragan2016}
{Banda-Barrag{\'a}n} W.~E.,  {Parkin} E.~R.,  {Federrath} C.,  {Crocker} R.~M.,
    {Bicknell} G.~V.,  2016, \mn@doi [\mnras] {10.1093/mnras/stv2405}, \href
  {https://ui.adsabs.harvard.edu/abs/2016MNRAS.455.1309B} {455, 1309}

\bibitem[\protect\citeauthoryear{{Banda-Barrag{\'a}n}, {Federrath}, {Crocker}
  \& {Bicknell}}{{Banda-Barrag{\'a}n} et~al.}{2018}]{Banda-Barragan2018}
{Banda-Barrag{\'a}n} W.~E.,  {Federrath} C.,  {Crocker} R.~M.,   {Bicknell}
  G.~V.,  2018, \mn@doi [\mnras] {10.1093/mnras/stx2541}, \href
  {https://ui.adsabs.harvard.edu/abs/2018MNRAS.473.3454B} {473, 3454}

\bibitem[\protect\citeauthoryear{{Banda-Barrag{\'a}n}, {Zertuche}, {Federrath},
  {Garc{\'\i}a Del Valle}, {Br{\"u}ggen}  \& {Wagner}}{{Banda-Barrag{\'a}n}
  et~al.}{2019}]{Banda2019}
{Banda-Barrag{\'a}n} W.~E.,  {Zertuche} F.~J.,  {Federrath} C.,  {Garc{\'\i}a
  Del Valle} J.,  {Br{\"u}ggen} M.,   {Wagner} A.~Y.,  2019, \mn@doi [\mnras]
  {10.1093/mnras/stz1040}, \href
  {https://ui.adsabs.harvard.edu/abs/2019MNRAS.486.4526B} {486, 4526}

\bibitem[\protect\citeauthoryear{{Braginskii}}{{Braginskii}}{1965}]{1965RvPP....1..205B}
{Braginskii} S.~I.,  1965, Reviews of Plasma Physics, \href
  {https://ui.adsabs.harvard.edu/\#abs/1965RvPP....1..205B} {1, 205}

\bibitem[\protect\citeauthoryear{{Br{\"u}ggen} \& {Scannapieco}}{{Br{\"u}ggen}
  \& {Scannapieco}}{2016}]{2016ApJ...822...31B}
{Br{\"u}ggen} M.,  {Scannapieco} E.,  2016, \mn@doi [\apj]
  {10.3847/0004-637X/822/1/31}, \href
  {https://ui.adsabs.harvard.edu/\#abs/2016ApJ...822...31B} {822, 31}

\bibitem[\protect\citeauthoryear{{Cen}}{{Cen}}{2013}]{2013ApJ...770..139C}
{Cen} R.,  2013, \mn@doi [\apj] {10.1088/0004-637X/770/2/139}, \href
  {https://ui.adsabs.harvard.edu/\#abs/2013ApJ...770..139C} {770, 139}

\bibitem[\protect\citeauthoryear{{Chen}, {Lanzetta}, {Webb}  \&
  {Barcons}}{{Chen} et~al.}{1998}]{1998ApJ...498...77C}
{Chen} H.-W.,  {Lanzetta} K.~M.,  {Webb} J.~K.,   {Barcons} X.,  1998, \mn@doi
  [\apj] {10.1086/305554}, \href
  {https://ui.adsabs.harvard.edu/\#abs/1998ApJ...498...77C} {498, 77}

\bibitem[\protect\citeauthoryear{{Churchill}, {Steidel}  \& {Vogt}}{{Churchill}
  et~al.}{1996}]{1996ApJ...471..164C}
{Churchill} C.~W.,  {Steidel} C.~C.,   {Vogt} S.~S.,  1996, \mn@doi [\apj]
  {10.1086/177960}, \href
  {https://ui.adsabs.harvard.edu/\#abs/1996ApJ...471..164C} {471, 164}

\bibitem[\protect\citeauthoryear{{Colbrook}, {Ma}, {Hopkins}  \&
  {Squire}}{{Colbrook} et~al.}{2017}]{colbrook:passive.scalar.scalings}
{Colbrook} M.~J.,  {Ma} X.,  {Hopkins} P.~F.,   {Squire} J.,  2017, \mn@doi
  [\mnras] {10.1093/mnras/stx261}, \href
  {http://adsabs.harvard.edu/abs/2017MNRAS.467.2421C} {467, 2421}

\bibitem[\protect\citeauthoryear{Cowie \& McKee}{Cowie \&
  McKee}{1977b}]{Cowie1977}
Cowie L.~L.,  McKee C.~F.,  1977b, \mn@doi [\apj] {10.1086/154911}, 211, 135

\bibitem[\protect\citeauthoryear{{Cowie} \& {McKee}}{{Cowie} \&
  {McKee}}{1977a}]{1977ApJ...211..135C}
{Cowie} L.~L.,  {McKee} C.~F.,  1977a, \mn@doi [\apj] {10.1086/154911}, \href
  {https://ui.adsabs.harvard.edu/\#abs/1977ApJ...211..135C} {211, 135}

\bibitem[\protect\citeauthoryear{{Dedner}, {Kemm}, {Kr{\"o}ner}, {Munz},
  {Schnitzer}  \& {Wesenberg}}{{Dedner} et~al.}{2002}]{Dedner2002}
{Dedner} A.,  {Kemm} F.,  {Kr{\"o}ner} D.,  {Munz} C.~D.,  {Schnitzer} T.,
  {Wesenberg} M.,  2002, \mn@doi [Journal of Computational Physics]
  {10.1006/jcph.2001.6961}, \href
  {https://ui.adsabs.harvard.edu/abs/2002JCoPh.175..645D} {175, 645}

\bibitem[\protect\citeauthoryear{{Dursi} \& {Pfrommer}}{{Dursi} \&
  {Pfrommer}}{2008}]{Dursi2008}
{Dursi} L.~J.,  {Pfrommer} C.,  2008, \mn@doi [\apj] {10.1086/529371}, \href
  {https://ui.adsabs.harvard.edu/abs/2008ApJ...677..993D} {677, 993}

\bibitem[\protect\citeauthoryear{{Faucher-Gigu{\`e}re}, {Lidz}, {Zaldarriaga}
  \& {Hernquist}}{{Faucher-Gigu{\`e}re} et~al.}{2009}]{2009ApJ...703.1416F}
{Faucher-Gigu{\`e}re} C.-A.,  {Lidz} A.,  {Zaldarriaga} M.,   {Hernquist} L.,
  2009, \mn@doi [\apj] {10.1088/0004-637X/703/2/1416}, \href
  {https://ui.adsabs.harvard.edu/\#abs/2009ApJ...703.1416F} {703, 1416}

\bibitem[\protect\citeauthoryear{{Faucher-Gigu{\`e}re}, {Hopkins},
  {Kere{\v{s}}}, {Muratov}, {Quataert}  \& {Murray}}{{Faucher-Gigu{\`e}re}
  et~al.}{2015}]{Faucher15}
{Faucher-Gigu{\`e}re} C.-A.,  {Hopkins} P.~F.,  {Kere{\v{s}}} D.,  {Muratov}
  A.~L.,  {Quataert} E.,   {Murray} N.,  2015, \mn@doi [\mnras]
  {10.1093/mnras/stv336}, \href
  {https://ui.adsabs.harvard.edu/abs/2015MNRAS.449..987F} {449, 987}

\bibitem[\protect\citeauthoryear{{Federrath} \& {Banerjee}}{{Federrath} \&
  {Banerjee}}{2015}]{Federrath15}
{Federrath} C.,  {Banerjee} S.,  2015, \mn@doi [\mnras] {10.1093/mnras/stv180},
  \href {https://ui.adsabs.harvard.edu/abs/2015MNRAS.448.3297F} {448, 3297}

\bibitem[\protect\citeauthoryear{{Ferland}, {Korista}, {Verner}, {Ferguson},
  {Kingdon}  \& {Verner}}{{Ferland} et~al.}{1998}]{1998PASP..110..761F}
{Ferland} G.~J.,  {Korista} K.~T.,  {Verner} D.~A.,  {Ferguson} J.~W.,
  {Kingdon} J.~B.,   {Verner} E.~M.,  1998, \mn@doi [\pasp] {10.1086/316190},
  \href {https://ui.adsabs.harvard.edu/abs/1998PASP..110..761F} {110, 761}

\bibitem[\protect\citeauthoryear{{Gronke} \& {Oh}}{{Gronke} \&
  {Oh}}{2018}]{Gronke18}
{Gronke} M.,  {Oh} S.~P.,  2018, \mn@doi [\mnras] {10.1093/mnrasl/sly131},
  \href {https://ui.adsabs.harvard.edu/\#abs/2018MNRAS.480L.111G} {480, L111}

\bibitem[\protect\citeauthoryear{{Gronke} \& {Oh}}{{Gronke} \&
  {Oh}}{2019}]{Gronke19}
{Gronke} M.,  {Oh} S.~P.,  2019, \mn@doi [\mnras] {10.1093/mnras/stz3332},
  \href {https://ui.adsabs.harvard.edu/abs/2019MNRAS.tmp.2995G} {p.~2995}

\bibitem[\protect\citeauthoryear{{Gr{\o}nnow}, {Tepper-Garc{\'\i}a}, {Bland
  -Hawthorn}  \& {McClure-Griffiths}}{{Gr{\o}nnow} et~al.}{2017}]{Gronnow2017}
{Gr{\o}nnow} A.,  {Tepper-Garc{\'\i}a} T.,  {Bland -Hawthorn} J.,
  {McClure-Griffiths} N.~M.,  2017, \mn@doi [\apj] {10.3847/1538-4357/aa7ed2},
  \href {https://ui.adsabs.harvard.edu/abs/2017ApJ...845...69G} {845, 69}

\bibitem[\protect\citeauthoryear{{Gr{\o}nnow}, {Tepper-Garc{\'\i}a}  \& {Bland
  -Hawthorn}}{{Gr{\o}nnow} et~al.}{2018}]{Gronnow2018}
{Gr{\o}nnow} A.,  {Tepper-Garc{\'\i}a} T.,   {Bland -Hawthorn} J.,  2018,
  \mn@doi [\apj] {10.3847/1538-4357/aada0e}, \href
  {https://ui.adsabs.harvard.edu/abs/2018ApJ...865...64G} {865, 64}

\bibitem[\protect\citeauthoryear{{Hopkins}}{{Hopkins}}{2015}]{2015MNRAS.450...53H}
{Hopkins} P.~F.,  2015, \mn@doi [\mnras] {10.1093/mnras/stv195}, \href
  {https://ui.adsabs.harvard.edu/\#abs/2015MNRAS.450...53H} {450, 53}

\bibitem[\protect\citeauthoryear{{Hopkins}}{{Hopkins}}{2016}]{divergence2016}
{Hopkins} P.~F.,  2016, \mn@doi [\mnras] {10.1093/mnras/stw1578}, \href
  {https://ui.adsabs.harvard.edu/abs/2016MNRAS.462..576H} {462, 576}

\bibitem[\protect\citeauthoryear{{Hopkins}}{{Hopkins}}{2017}]{2017MNRAS.466.3387H}
{Hopkins} P.~F.,  2017, \mn@doi [\mnras] {10.1093/mnras/stw3306}, \href
  {https://ui.adsabs.harvard.edu/\#abs/2017MNRAS.466.3387H} {466, 3387}

\bibitem[\protect\citeauthoryear{{Hopkins} \& {Raives}}{{Hopkins} \&
  {Raives}}{2016}]{2016MNRAS.455...51H}
{Hopkins} P.~F.,  {Raives} M.~J.,  2016, \mn@doi [\mnras]
  {10.1093/mnras/stv2180}, \href
  {https://ui.adsabs.harvard.edu/\#abs/2016MNRAS.455...51H} {455, 51}

\bibitem[\protect\citeauthoryear{{Hopkins} et~al.,}{{Hopkins}
  et~al.}{2018}]{2018MNRAS.480..800H}
{Hopkins} P.~F.,  et~al., 2018, \mn@doi [\mnras] {10.1093/mnras/sty1690}, \href
  {https://ui.adsabs.harvard.edu/\#abs/2018MNRAS.480..800H} {480, 800}

\bibitem[\protect\citeauthoryear{{Hopkins} et~al.,}{{Hopkins}
  et~al.}{2019}]{Hopkins19}
{Hopkins} P.~F.,  et~al., 2019, \mn@doi [\mnras] {10.1093/mnras/stz3321}, \href
  {https://ui.adsabs.harvard.edu/abs/2019MNRAS.tmp.2993H} {p.~2993}

\bibitem[\protect\citeauthoryear{{Johnson}, {Chen}, {Mulchaey}, {Schaye}  \&
  {Straka}}{{Johnson} et~al.}{2017}]{2017ApJ...850L..10J}
{Johnson} S.~D.,  {Chen} H.-W.,  {Mulchaey} J.~S.,  {Schaye} J.,   {Straka}
  L.~A.,  2017, \mn@doi [\apj] {10.3847/2041-8213/aa9370}, \href
  {https://ui.adsabs.harvard.edu/\#abs/2017ApJ...850L..10J} {850, L10}

\bibitem[\protect\citeauthoryear{{Jones}, {Ryu}  \& {Tregillis}}{{Jones}
  et~al.}{1996}]{1996ApJ...473..365J}
{Jones} T.~W.,  {Ryu} D.,   {Tregillis} I.~L.,  1996, \mn@doi [\apj]
  {10.1086/178151}, \href
  {https://ui.adsabs.harvard.edu/\#abs/1996ApJ...473..365J} {473, 365}

\bibitem[\protect\citeauthoryear{{Kawazura}, {Barnes}  \&
  {Schekochihin}}{{Kawazura} et~al.}{2019}]{Kawazura2019}
{Kawazura} Y.,  {Barnes} M.,   {Schekochihin} A.~A.,  2019, \mn@doi
  [Proceedings of the National Academy of Science] {10.1073/pnas.1812491116},
  \href {https://ui.adsabs.harvard.edu/abs/2019PNAS..116..771K} {116, 771}

\bibitem[\protect\citeauthoryear{{Kere{\v{s}}} \& {Hernquist}}{{Kere{\v{s}}} \&
  {Hernquist}}{2009}]{2009ApJ...700L...1K}
{Kere{\v{s}}} D.,  {Hernquist} L.,  2009, \mn@doi [\apj]
  {10.1088/0004-637X/700/1/L1}, \href
  {https://ui.adsabs.harvard.edu/\#abs/2009ApJ...700L...1K} {700, L1}

\bibitem[\protect\citeauthoryear{{Klein}, {McKee}  \& {Colella}}{{Klein}
  et~al.}{1994}]{Klein1994}
{Klein} R.~I.,  {McKee} C.~F.,   {Colella} P.,  1994, \mn@doi [\apj]
  {10.1086/173554}, \href
  {https://ui.adsabs.harvard.edu/abs/1994ApJ...420..213K} {420, 213}

\bibitem[\protect\citeauthoryear{{Komarov}, {Churazov}, {Kunz}  \&
  {Schekochihin}}{{Komarov} et~al.}{2016}]{Komarov2016}
{Komarov} S.~V.,  {Churazov} E.~M.,  {Kunz} M.~W.,   {Schekochihin} A.~A.,
  2016, \mn@doi [\mnras] {10.1093/mnras/stw963}, \href
  {https://ui.adsabs.harvard.edu/\#abs/2016MNRAS.460..467K} {460, 467}

\bibitem[\protect\citeauthoryear{{K{\"o}rtgen}, {Federrath}  \&
  {Banerjee}}{{K{\"o}rtgen} et~al.}{2019}]{Federrath19}
{K{\"o}rtgen} B.,  {Federrath} C.,   {Banerjee} R.,  2019, \mn@doi [\mnras]
  {10.1093/mnras/sty3071}, \href
  {https://ui.adsabs.harvard.edu/abs/2019MNRAS.482.5233K} {482, 5233}

\bibitem[\protect\citeauthoryear{{Krumholz} \& {Gnedin}}{{Krumholz} \&
  {Gnedin}}{2011}]{2011ApJ...729...36K}
{Krumholz} M.~R.,  {Gnedin} N.~Y.,  2011, \mn@doi [\apj]
  {10.1088/0004-637X/729/1/36}, \href
  {https://ui.adsabs.harvard.edu/\#abs/2011ApJ...729...36K} {729, 36}

\bibitem[\protect\citeauthoryear{Kulsrud}{Kulsrud}{1983}]{Kulsrud1983}
Kulsrud R.~M.,  1983, in Sagdeev R.~N.,  Rosenbluth M.~N.,  eds, , {Handbook of
  Plasma Physics}.
Princeton University

\bibitem[\protect\citeauthoryear{Kunz, Schekochihin  \& Stone}{Kunz
  et~al.}{2014}]{Kunz2014}
Kunz M.~W.,  Schekochihin A.~A.,   Stone J.~M.,  2014, \mn@doi [\prl]
  {10.1103/physrevlett.112.205003}, 112, 205003

\bibitem[\protect\citeauthoryear{{Li}, {Frank}  \& {Blackman}}{{Li}
  et~al.}{2013}]{Li2013}
{Li} S.,  {Frank} A.,   {Blackman} E.~G.,  2013, \mn@doi [\apj]
  {10.1088/0004-637X/774/2/133}, \href
  {https://ui.adsabs.harvard.edu/abs/2013ApJ...774..133L} {774, 133}

\bibitem[\protect\citeauthoryear{{Li}, {Frank}  \& {Blackman}}{{Li}
  et~al.}{2014}]{Li2014}
{Li} S.,  {Frank} A.,   {Blackman} E.~G.,  2014, \mn@doi [\mnras]
  {10.1093/mnras/stu1571}, \href
  {https://ui.adsabs.harvard.edu/abs/2014MNRAS.444.2884L} {444, 2884}

\bibitem[\protect\citeauthoryear{{Liang} \& {Remming}}{{Liang} \&
  {Remming}}{2018}]{2018arXiv180610688L}
{Liang} C.~J.,  {Remming} I.~S.,  2018, arXiv e-prints, \href
  {https://ui.adsabs.harvard.edu/\#abs/2018arXiv180610688L} {p.
  arXiv:1806.10688}

\bibitem[\protect\citeauthoryear{{Liang}, {Kravtsov}  \& {Agertz}}{{Liang}
  et~al.}{2016}]{2016MNRAS.458.1164L}
{Liang} C.~J.,  {Kravtsov} A.~V.,   {Agertz} O.,  2016, \mn@doi [\mnras]
  {10.1093/mnras/stw375}, \href
  {https://ui.adsabs.harvard.edu/\#abs/2016MNRAS.458.1164L} {458, 1164}

\bibitem[\protect\citeauthoryear{{Mac Low}, {McKee}, {Klein}, {Stone}  \&
  {Norman}}{{Mac Low} et~al.}{1994}]{1994ApJ...433..757M}
{Mac Low} M.-M.,  {McKee} C.~F.,  {Klein} R.~I.,  {Stone} J.~M.,   {Norman}
  M.~L.,  1994, \mn@doi [\apj] {10.1086/174685}, \href
  {https://ui.adsabs.harvard.edu/\#abs/1994ApJ...433..757M} {433, 757}

\bibitem[\protect\citeauthoryear{{Markevitch} \& {Vikhlinin}}{{Markevitch} \&
  {Vikhlinin}}{2007}]{2007PhR...443....1M}
{Markevitch} M.,  {Vikhlinin} A.,  2007, \mn@doi [\physrep]
  {10.1016/j.physrep.2007.01.001}, \href
  {https://ui.adsabs.harvard.edu/\#abs/2007PhR...443....1M} {443, 1}

\bibitem[\protect\citeauthoryear{{Martin-Alvarez}, {Devriendt}, {Slyz}  \&
  {Teyssier}}{{Martin-Alvarez} et~al.}{2018}]{Martin2018}
{Martin-Alvarez} S.,  {Devriendt} J.,  {Slyz} A.,   {Teyssier} R.,  2018,
  \mn@doi [\mnras] {10.1093/mnras/sty1623}, \href
  {https://ui.adsabs.harvard.edu/abs/2018MNRAS.479.3343M} {479, 3343}

\bibitem[\protect\citeauthoryear{{McCourt}, {O'Leary}, {Madigan}  \&
  {Quataert}}{{McCourt} et~al.}{2015}]{McCourt15}
{McCourt} M.,  {O'Leary} R.~M.,  {Madigan} A.-M.,   {Quataert} E.,  2015,
  \mn@doi [\mnras] {10.1093/mnras/stv355}, \href
  {https://ui.adsabs.harvard.edu/abs/2015MNRAS.449....2M} {449, 2}

\bibitem[\protect\citeauthoryear{{McCourt}, {Oh}, {O'Leary}  \&
  {Madigan}}{{McCourt} et~al.}{2018}]{2018MNRAS.473.5407M}
{McCourt} M.,  {Oh} S.~P.,  {O'Leary} R.,   {Madigan} A.-M.,  2018, \mn@doi
  [\mnras] {10.1093/mnras/stx2687}, \href
  {https://ui.adsabs.harvard.edu/\#abs/2018MNRAS.473.5407M} {473, 5407}

\bibitem[\protect\citeauthoryear{{McKee} \& {Cowie}}{{McKee} \&
  {Cowie}}{1975}]{1975ApJ...195..715M}
{McKee} C.~F.,  {Cowie} L.~L.,  1975, \mn@doi [\apj] {10.1086/153373}, \href
  {https://ui.adsabs.harvard.edu/\#abs/1975ApJ...195..715M} {195, 715}

\bibitem[\protect\citeauthoryear{{McKee} \& {Cowie}}{{McKee} \&
  {Cowie}}{1977}]{1977ApJ...215..213M}
{McKee} C.~F.,  {Cowie} L.~L.,  1977, \mn@doi [\apj] {10.1086/155350}, \href
  {https://ui.adsabs.harvard.edu/\#abs/1977ApJ...215..213M} {215, 213}

\bibitem[\protect\citeauthoryear{{Miniati}, {Jones}  \& {Ryu}}{{Miniati}
  et~al.}{1999}]{1999ApJ...517..242M}
{Miniati} F.,  {Jones} T.~W.,   {Ryu} D.,  1999, \mn@doi [\apj]
  {10.1086/307162}, \href
  {https://ui.adsabs.harvard.edu/\#abs/1999ApJ...517..242M} {517, 242}

\bibitem[\protect\citeauthoryear{{Mouschovias}}{{Mouschovias}}{1976a}]{1976ApJ...206..753M}
{Mouschovias} T.~C.,  1976a, \mn@doi [\apj] {10.1086/154436}, \href
  {https://ui.adsabs.harvard.edu/abs/1976ApJ...206..753M} {206, 753}

\bibitem[\protect\citeauthoryear{{Mouschovias}}{{Mouschovias}}{1976b}]{1976ApJ...207..141M}
{Mouschovias} T.~C.,  1976b, \mn@doi [\apj] {10.1086/154478}, \href
  {https://ui.adsabs.harvard.edu/abs/1976ApJ...207..141M} {207, 141}

\bibitem[\protect\citeauthoryear{{Orr} et~al.,}{{Orr}
  et~al.}{2018}]{2018MNRAS.478.3653O}
{Orr} M.~E.,  et~al., 2018, \mn@doi [\mnras] {10.1093/mnras/sty1241}, \href
  {https://ui.adsabs.harvard.edu/\#abs/2018MNRAS.478.3653O} {478, 3653}

\bibitem[\protect\citeauthoryear{{Price} \& {Monaghan}}{{Price} \&
  {Monaghan}}{2007}]{2007MNRAS.374.1347P}
{Price} D.~J.,  {Monaghan} J.~J.,  2007, \mn@doi [\mnras]
  {10.1111/j.1365-2966.2006.11241.x}, \href
  {https://ui.adsabs.harvard.edu/\#abs/2007MNRAS.374.1347P} {374, 1347}

\bibitem[\protect\citeauthoryear{{Prochaska}, {Lau}  \& {Hennawi}}{{Prochaska}
  et~al.}{2014}]{2014ApJ...796..140P}
{Prochaska} J.~X.,  {Lau} M.~W.,   {Hennawi} J.~F.,  2014, \mn@doi [\apj]
  {10.1088/0004-637X/796/2/140}, \href
  {https://ui.adsabs.harvard.edu/\#abs/2014ApJ...796..140P} {796, 140}

\bibitem[\protect\citeauthoryear{{Rahmati}, {Pawlik}, {Rai{\v{c}}evi{\'c}}  \&
  {Schaye}}{{Rahmati} et~al.}{2013}]{2013MNRAS.430.2427R}
{Rahmati} A.,  {Pawlik} A.~H.,  {Rai{\v{c}}evi{\'c}} M.,   {Schaye} J.,  2013,
  \mn@doi [\mnras] {10.1093/mnras/stt066}, \href
  {https://ui.adsabs.harvard.edu/\#abs/2013MNRAS.430.2427R} {430, 2427}

\bibitem[\protect\citeauthoryear{{Richter}, {Paerels}  \& {Kaastra}}{{Richter}
  et~al.}{2008}]{2008SSRv..134...25R}
{Richter} P.,  {Paerels} F.~B.~S.,   {Kaastra} J.~S.,  2008, \mn@doi [\ssr]
  {10.1007/s11214-008-9325-4}, \href
  {https://ui.adsabs.harvard.edu/\#abs/2008SSRv..134...25R} {134, 25}

\bibitem[\protect\citeauthoryear{{Robertson} \& {Kravtsov}}{{Robertson} \&
  {Kravtsov}}{2008}]{2008ApJ...680.1083R}
{Robertson} B.~E.,  {Kravtsov} A.~V.,  2008, \mn@doi [\apj] {10.1086/587796},
  \href {https://ui.adsabs.harvard.edu/\#abs/2008ApJ...680.1083R} {680, 1083}

\bibitem[\protect\citeauthoryear{{Sarazin}}{{Sarazin}}{1988}]{1988xrec.book.....S}
{Sarazin} C.~L.,  1988, {X-ray emission from clusters of galaxies}

\bibitem[\protect\citeauthoryear{{Savage}, {Lehner}  \& {Narayanan}}{{Savage}
  et~al.}{2011}]{2011ApJ...743..180S}
{Savage} B.~D.,  {Lehner} N.,   {Narayanan} A.,  2011, \mn@doi [\apj]
  {10.1088/0004-637X/743/2/180}, \href
  {https://ui.adsabs.harvard.edu/\#abs/2011ApJ...743..180S} {743, 180}

\bibitem[\protect\citeauthoryear{{Scannapieco} \& {Br{\"u}ggen}}{{Scannapieco}
  \& {Br{\"u}ggen}}{2015}]{2015ApJ...805..158S}
{Scannapieco} E.,  {Br{\"u}ggen} M.,  2015, \mn@doi [\apj]
  {10.1088/0004-637X/805/2/158}, \href
  {https://ui.adsabs.harvard.edu/\#abs/2015ApJ...805..158S} {805, 158}

\bibitem[\protect\citeauthoryear{{Schneider} \& {Robertson}}{{Schneider} \&
  {Robertson}}{2015}]{2015ApJS..217...24S}
{Schneider} E.~E.,  {Robertson} B.~E.,  2015, \mn@doi [\apjs]
  {10.1088/0067-0049/217/2/24}, \href
  {https://ui.adsabs.harvard.edu/abs/2015ApJS..217...24S} {217, 24}

\bibitem[\protect\citeauthoryear{{Shin}, {Stone}  \& {Snyder}}{{Shin}
  et~al.}{2008}]{2008ApJ...680..336S}
{Shin} M.-S.,  {Stone} J.~M.,   {Snyder} G.~F.,  2008, \mn@doi [\apj]
  {10.1086/587775}, \href
  {https://ui.adsabs.harvard.edu/\#abs/2008ApJ...680..336S} {680, 336}

\bibitem[\protect\citeauthoryear{{Sparre}, {Pfrommer}  \&
  {Vogelsberger}}{{Sparre} et~al.}{2019}]{Sparre2019}
{Sparre} M.,  {Pfrommer} C.,   {Vogelsberger} M.,  2019, \mn@doi [\mnras]
  {10.1093/mnras/sty3063}, \href
  {https://ui.adsabs.harvard.edu/abs/2019MNRAS.482.5401S} {482, 5401}

\bibitem[\protect\citeauthoryear{{Spitzer} \& {H{\"a}rm}}{{Spitzer} \&
  {H{\"a}rm}}{1953}]{1953PhRv...89..977S}
{Spitzer} L.,  {H{\"a}rm} R.,  1953, \mn@doi [Physical Review]
  {10.1103/PhysRev.89.977}, \href
  {https://ui.adsabs.harvard.edu/\#abs/1953PhRv...89..977S} {89, 977}

\bibitem[\protect\citeauthoryear{{Springel}}{{Springel}}{2005}]{2005MNRAS.364.1105S}
{Springel} V.,  2005, \mn@doi [\mnras] {10.1111/j.1365-2966.2005.09655.x},
  \href {https://ui.adsabs.harvard.edu/\#abs/2005MNRAS.364.1105S} {364, 1105}

\bibitem[\protect\citeauthoryear{{Squire}, {Schekochihin}, {Quataert}  \&
  {Kunz}}{{Squire} et~al.}{2019}]{Squire2019}
{Squire} J.,  {Schekochihin} A.~A.,  {Quataert} E.,   {Kunz} M.~W.,  2019,
  \mn@doi [Journal of Plasma Physics] {10.1017/S0022377819000114}, \href
  {https://ui.adsabs.harvard.edu/abs/2019JPlPh..85a9014S} {85, 905850114}

\bibitem[\protect\citeauthoryear{{Steidel}, {Erb}, {Shapley}, {Pettini},
  {Reddy}, {Bogosavljevi{\'c}}, {Rudie}  \& {Rakic}}{{Steidel}
  et~al.}{2010}]{2010ApJ...717..289S}
{Steidel} C.~C.,  {Erb} D.~K.,  {Shapley} A.~E.,  {Pettini} M.,  {Reddy} N.,
  {Bogosavljevi{\'c}} M.,  {Rudie} G.~C.,   {Rakic} O.,  2010, \mn@doi [\apj]
  {10.1088/0004-637X/717/1/289}, \href
  {https://ui.adsabs.harvard.edu/\#abs/2010ApJ...717..289S} {717, 289}

\bibitem[\protect\citeauthoryear{{Stocke}, {Penton}, {Danforth}, {Shull},
  {Tumlinson}  \& {McLin}}{{Stocke} et~al.}{2006}]{2006ApJ...641..217S}
{Stocke} J.~T.,  {Penton} S.~V.,  {Danforth} C.~W.,  {Shull} J.~M.,
  {Tumlinson} J.,   {McLin} K.~M.,  2006, \mn@doi [\apj] {10.1086/500386},
  \href {https://ui.adsabs.harvard.edu/\#abs/2006ApJ...641..217S} {641, 217}

\bibitem[\protect\citeauthoryear{{Su}, {Hopkins}, {Hayward},
  {Faucher-Gigu{\`e}re}, {Kere{\v{s}}}, {Ma}  \& {Robles}}{{Su}
  et~al.}{2017}]{2017MNRAS.471..144S}
{Su} K.-Y.,  {Hopkins} P.~F.,  {Hayward} C.~C.,  {Faucher-Gigu{\`e}re} C.-A.,
  {Kere{\v{s}}} D.,  {Ma} X.,   {Robles} V.~H.,  2017, \mn@doi [\mnras]
  {10.1093/mnras/stx1463}, \href
  {https://ui.adsabs.harvard.edu/\#abs/2017MNRAS.471..144S} {471, 144}

\bibitem[\protect\citeauthoryear{{Thompson}, {Quataert}, {Zhang}  \&
  {Weinberg}}{{Thompson} et~al.}{2016}]{2016MNRAS.455.1830T}
{Thompson} T.~A.,  {Quataert} E.,  {Zhang} D.,   {Weinberg} D.~H.,  2016,
  \mn@doi [\mnras] {10.1093/mnras/stv2428}, \href
  {https://ui.adsabs.harvard.edu/\#abs/2016MNRAS.455.1830T} {455, 1830}

\bibitem[\protect\citeauthoryear{{Tumlinson}, {Peeples}  \& {Werk}}{{Tumlinson}
  et~al.}{2017}]{2017ARA&A..55..389T}
{Tumlinson} J.,  {Peeples} M.~S.,   {Werk} J.~K.,  2017, \mn@doi [\araa]
  {10.1146/annurev-astro-091916-055240}, \href
  {https://ui.adsabs.harvard.edu/abs/2017ARA&A..55..389T} {55, 389}

\bibitem[\protect\citeauthoryear{{Werk} et~al.,}{{Werk}
  et~al.}{2014}]{2014ApJ...792....8W}
{Werk} J.~K.,  et~al., 2014, \mn@doi [\apj] {10.1088/0004-637X/792/1/8}, \href
  {https://ui.adsabs.harvard.edu/\#abs/2014ApJ...792....8W} {792, 8}

\bibitem[\protect\citeauthoryear{{Werk} et~al.,}{{Werk}
  et~al.}{2016}]{2016ApJ...833...54W}
{Werk} J.~K.,  et~al., 2016, \mn@doi [\apj] {10.3847/1538-4357/833/1/54}, \href
  {https://ui.adsabs.harvard.edu/\#abs/2016ApJ...833...54W} {833, 54}

\bibitem[\protect\citeauthoryear{{Yao}, {Wang}, {Penton}, {Tripp}, {Shull}  \&
  {Stocke}}{{Yao} et~al.}{2010}]{2010ApJ...716.1514Y}
{Yao} Y.,  {Wang} Q.~D.,  {Penton} S.~V.,  {Tripp} T.~M.,  {Shull} J.~M.,
  {Stocke} J.~T.,  2010, \mn@doi [\apj] {10.1088/0004-637X/716/2/1514}, \href
  {https://ui.adsabs.harvard.edu/\#abs/2010ApJ...716.1514Y} {716, 1514}

\makeatother
\end{thebibliography}
\begin{appendix}

\section{Convergence Tests}\label{sec:convergence}

\begin{figure}
\includegraphics[width=0.49\textwidth]{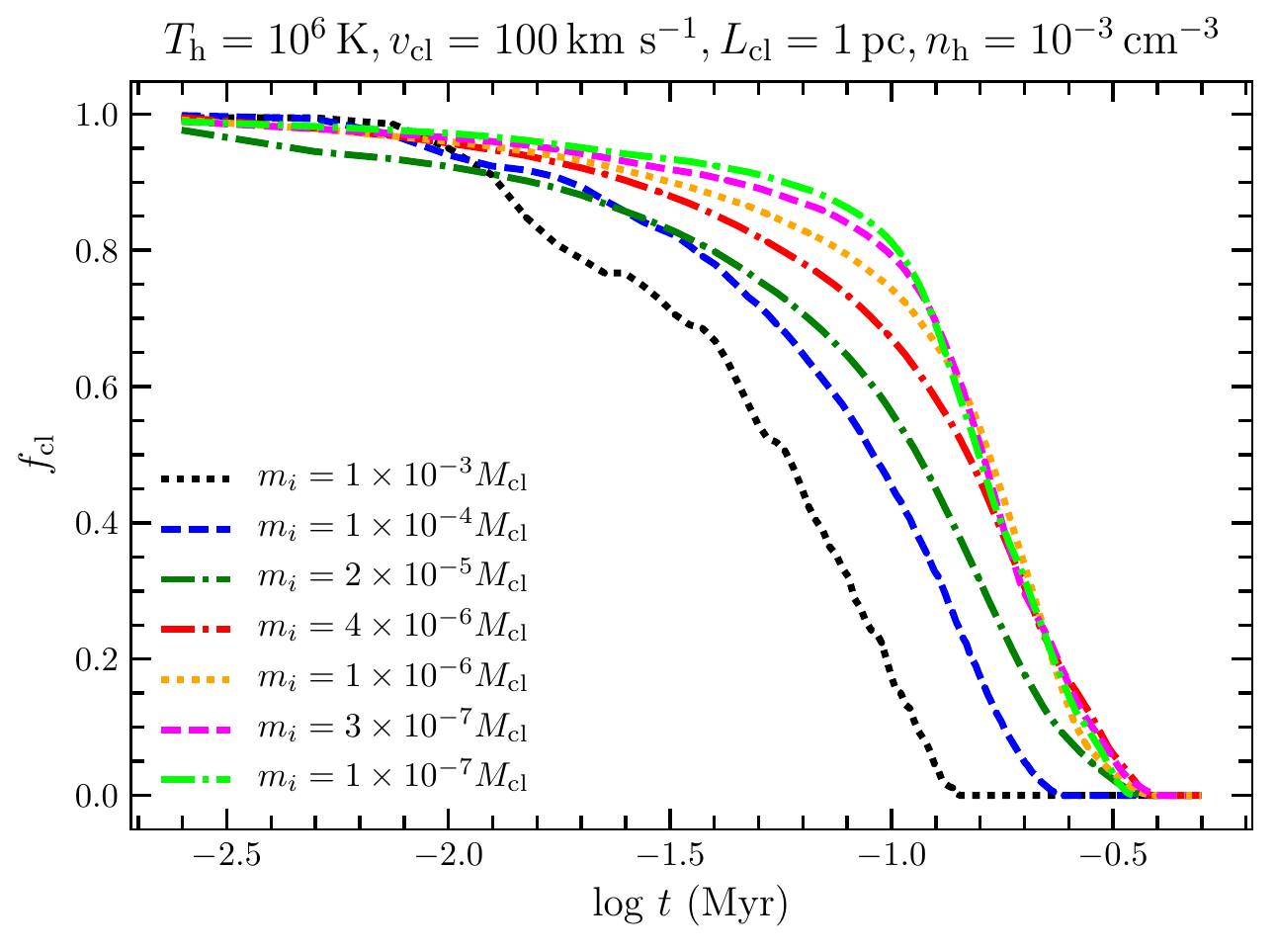}
    \vspace{-0.25cm}
    \caption{Evolution of the normalized cloud mass, $f_{\rm cl}$ (defined in Figure \ref{fig:mcloud_definition}) versus time, for one representative initial condition in the ``classical cloud destruction'' regime ($T_{\rm h}$ = 10$^6$\,K, $v_{\rm cl}$ = 100\,${\rm km\,s^{-1}}$, $n_{\rm h}$ = 10$^{-3}\,{\rm cm^{-3}}$, $L_{\rm cl}$ = 1\,pc) with our default physics set simulated at seven different mass resolution ($m_{i}$) levels, as labeled. The resulting cloud lifetime is remarkably robust to resolution, changing by $<10\%$ from $m_{i}/M_{\rm cl} \sim 10^{-5}-10^{-7}$ and by a factor of $<2$ ($<3$) even at resolutions $m_{i}/M_{\rm cl} \sim 10^{-4}$ ($\sim 10^{-3}$). Recall our default resolution in the main text is $m_{i}/M_{\rm cl} \sim 10^{-6}$. The small change in behavior at early times and high resolution (with a longer ``delay'' until destruction begins) owes to better tracking of small, high-density ``features'' (e.g,.\ Kelvin-Helmholtz whorls) which remain locally high density even as mixing begins.
    \label{fig:resolution}}
\end{figure}

We have verified that our results are robust to numerical resolution ($m_{i} \sim 10^{-7}-10^{-3}\,M_{\rm cl}$, or equivalently, $ \sim\,$134 -- 6 cells per $R_{\rm cl}$) via a variety of tests. For at least one cloud in every ``regime'' shown in Figure \ref{fig:regime}, we have re-run the same initial conditions at three resolution levels (our default, and one and two orders-of-magnitude lower resolution). In all cases we confirm that the measured cloud lifetime is robust to better than a factor of $\sim2$ (although the cloud lifetimes do become systematically shorter at low resolution, as expected owing to numerical mixing). We have also randomly selected ten clouds in the ``classical cloud destruction'' regime to simulate at both lower and higher resolutions (a factor of $\sim 8$ change): we find the lifetimes change by a factor of $<1.5$ in these cases. In Figure \ref{fig:resolution}, we show one fiducial cloud, for which we simulate at seven different resolution levels. The agreement in cloud lifetime is excellent at order-of-magnitude higher and lower resolutions, compared to our default choice in the main text, which lends confidence to our conclusion that our key results are not strongly sensitive to numerical resolution.

\end{appendix}
\end{document}